\documentclass[%
 aip,
 amsmath,amssymb,
 reprint,%
]{revtex4-1}

\usepackage{graphicx}
\usepackage{dcolumn}
\usepackage{bm}

\usepackage[utf8]{inputenc}
\usepackage[T1]{fontenc}
\usepackage{mathptmx}

\graphicspath{{imglocal/} {./}} 

\begin{document}

\preprint{AIP/123-QED}

\title[Memristive Neuromorphic Systems]{A recipe for creating ideal hybrid memristive-CMOS neuromorphic computing systems}

\author{E. Chicca}
 \email{chicca@cit-ec.uni-bielefeld.de}
 \affiliation{Faculty of Technology and Cognitive Interaction Technology - Center of Excellence (CITEC), Bielefeld University, Bielefeld, Germany}
\author{G. Indiveri}%
 \email{giacomo@ini.uzh.ch}
\affiliation{Institute of Neuroinformatics, University of Zurich and ETH Zurich, Switzerland
}%

\date{\today}

\begin{abstract}
The development of memristive device technologies has reached a level of maturity to enable the design of complex and large-scale hybrid memristive-CMOS neural processing systems. These systems offer promising solutions for implementing novel in-memory computing architectures for machine learning and data analysis problems. We argue that they are also ideal building blocks for the integration in neuromorphic electronic circuits suitable for ultra-low power brain-inspired sensory processing systems, therefore leading to the  innovative solutions for always-on edge-computing and Internet-of-Things (IoT) applications. Here we present a recipe for creating such systems based on design strategies and computing principles inspired by those used in mammalian brains. We enumerate the specifications and properties of memristive devices required to support always-on learning in neuromorphic computing system and to minimize their power consumption. Finally, we discuss in what cases such neuromorphic systems can complement conventional processing ones and highlight the  importance of exploiting the physics of both the memristive devices and of the CMOS circuits interfaced to them.
\end{abstract}

\maketitle


Neuromorphic computing has recently received considerable attention as a discipline that can offer promising technological solutions for implementing power- and size-efficient sensory-processing, learning, and Artificial Intelligence (AI) applications~\cite{Neftci18,Thakur_etal18,Li_etal18c,van-De-Burgt_etal18,Burr_etal17}, especially in cases in which the computing system has to operate autonomously ``at the edge'', i.e., without having to connect to powerful (but power hungry) server farms in the ``cloud''.
The term ``neuromorphic'' was originally coined in the early 90's by Carver Mead to refer to mixed signal analog/digital Very Large Scale Integration (VLSI) computing systems based on the organizing principles used by the biological nervous systems~\cite{Mead90}. 
In that context, ``neuromorphic engineering'' emerged as an interdisciplinary research field deeply rooted in biology that focused on building electronic neural processing systems by exploiting the physics of silicon to directly ``emulate'' the bio-physics of real neurons and synapses. 
More recently the definition of the term ``neuromorphic'' has been extended in two additional directions: on one hand to describe more generic spike-based processing systems engineered to ``simulate'' spiking neural networks for the exploration of large-scale computational neuroscience models~\cite{Furber_etal14,Akopyan_etal15,Davies_etal18}; and on the other hand to describe dedicated electronic neural architectures that make use of both electronic Complementary Metal-Oxide Semiconductor (CMOS) circuits and memristive devices to implement neuron and synapse circuits~\cite{Ielmini_Waser15,Boybat_etal18}.

Another recent and very promising trend in developing dedicated hardware architectures for building accelerated simulators of artificial neural networks is related to the field of machine learning and AI~\cite{LeCun_etal15,Schmidhuber15}.
The types of neural networks being proposed within this context are only loosely inspired by biology, are aimed at high accuracy pattern recognition based on large data-sets, and require large amounts of memory for storing network states and parameters.
While this approach is producing amazing results in a wide range of application areas, the computing systems used to simulate these networks use significant amount of compute resources and power, especially for the training phase: the learning algorithms rely on high precision digital representations for calculating high accuracy gradients, and they typically require the storage (and transfer from peripheral memory to central processing areas) of very large data-sets. 
Furthermore, they often separate the training from the inference phase, dismissing the ability to adapt to novel stimuli and changing environmental conditions, typical of biological systems. 

\begin{figure}
  \centering
  \includegraphics[width=0.45\textwidth]{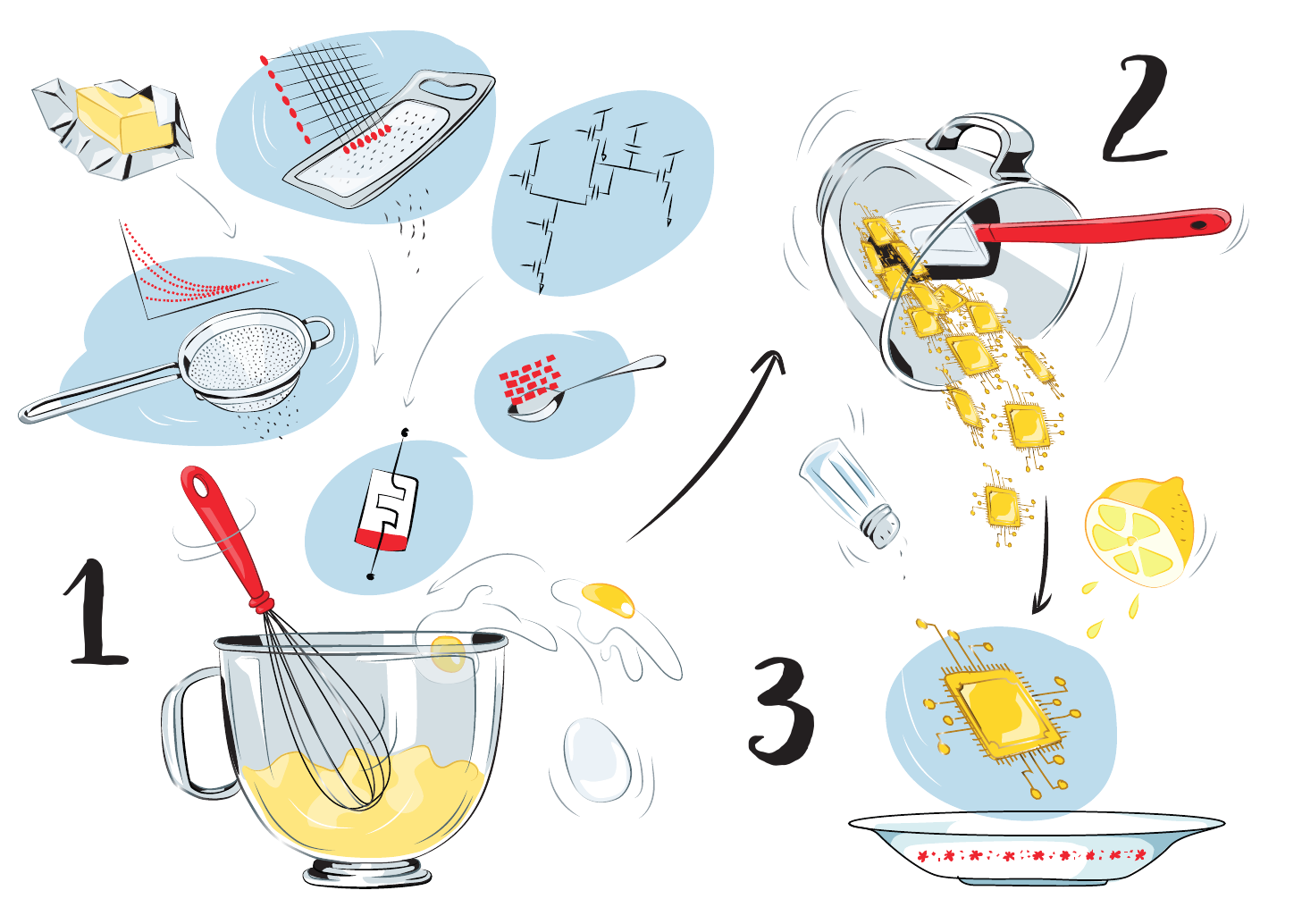}
  \caption{The ideal memristive neuromorphic computing system requires the right mix of CMOS circuits and memristive devices, as well as the proper use of spatial resources and temporal dynamics, that need to be well matched to the system's signal-processing applications and use-cases.}
  \label{fig:recipe}
\end{figure}

While there are examples of hybrid memristive-CMOS hardware architectures being developed to provide support for AI deep network accelerators~\cite{Sebastian_etal19,Boybat_etal18,Ambrogio_etal18,Burr_etal17}, it is important to clarify that many of the hybrid memristive-CMOS neuromorphic circuits proposed in the literature~\cite{Dai_etal19,Covi_etal16,Jo_etal10,Yang_etal17,Berdan_etal16} as well as the original neuromorphic approach of emulating biological neural systems proposed by Mead, are distinct and complementary to the machine learning one. While the machine learning approach is based on software algorithms developed to minimize the recognition error in very specific pattern recognition tasks, the original neuromorphic approach is based on brain-inspired electronic circuits and hardware architectures designed for reproducing the function of cortical and biological neural circuits~\cite{Chicca_etal14}. As a consequence, this approach aims at understanding how to build robust and low-power neural processing systems using inhomogeneous and highly variable components, fault-tolerant massively parallel arrays of computational elements, and in-memory computing (non von Neumann) information processing architectures~\cite{Indiveri_Liu15}.
In the following, when discussing about ``hybrid CMOS-memristive neuromorphic computing systems'', we will refer to this specific approach.



Our \emph{recipe} (Fig.~\ref{fig:recipe}) for optimally building neuromorphic systems by co-integrating memristive devices with CMOS circuits is based on the following considerations.

\paragraph{Lay out the ingredients in parallel on the worktop.}
To minimize power consumption and maximize robustness to variability, it is important to use physically distinct instantiations of neuron and synapse circuits, distributed across the silicon substrate~\cite{Indiveri_Sandamirskaya19}. This strategy is very different from the one used to build classical computing systems based on the von Neumann architecture. In classical processors there is a single or a small number of computing blocks that are time-multiplexed at very high clock rates to execute calculations, or to simulate many ``parallel'' neural processes~\cite{Furber_etal14,Merolla_etal14a,Davies_etal18}. The continuous transfer of data between memory and the time-multiplexed processing unit(s) required to carry out computation is limited by the infamous von Neumann bottleneck~\cite{Backus78}, and is the major cause of high energy consumption. In contrast, the amazing energy efficiency of biological systems, and of the neuromorphic ones that emulate them, arises from the in-memory computing nature of their architectures: there are multiple instances of neuron and synapse elements that carry out computation and at the same time store the network state. The disadvantage of having distributed state-full neuron and synapse circuits is that it can require significant amount of silicon real-estate for integrating all their memory structures (e.g., see the $4.3\,$cm$^2$ IBM TrueNorth chip~\cite{Merolla_etal14a}). However,  the progress in CMOS fabrication technologies, the emergence of monolithic 3D integration technologies, and the possibility to co-integrate nano-scale memristive devices with mixed-signal analog/digital CMOS circuits in advanced node processes can substantially mitigate this problem\cite{Indiveri_etal13}.

\paragraph{Take your time.}
By eliminating the need to use time-multiplexed processing elements, these neuromorphic processing architectures can be designed to run in real physical time (time represents itself) as it happens in real biological neural networks. 
This is a radical departure from the classical way of implementing computation, that has decoupled computer simulation time from physical time since the very early designs of both computing systems and artificial neural networks~\cite{Von-Neumann93,McCulloch_Pitts43}. 
For sensory-motor processing systems and edge-computing applications that need to measure and process natural signals, this is a tremendous advantage. Allowing time to represent itself removes the need of complicated clock or synchronizing structures that would otherwise be required to track the passage of simulated time. 
All computing elements in such neuromorphic systems are then coupled through the common variable of real-time (e.g., for implementing binding by synchronization\cite{Shadlen_Movshon99}). 
To build sensory-processing systems that are best tuned to the signals they are required to process (or that can learn to extract information from them), it is necessary to use neural processing and learning circuits that have the same time-constants and dynamics of their input signals (e.g., to create ``matched-filter'' that can naturally resonate with their inputs). 
In the case of natural signals typically processed by humans, such as voice or gestures, these time constants should range from milliseconds to minutes or longer. 
These time constants are extremely long, compared to the typical processing rates of digital circuits.
This allows neuromorphic systems to reduce power consumption even more and to have very large bandwidths for seamlessly transmitting signals across the network and via I/O pathways in shared buses~\cite{Boahen98,Moradi_etal18}.
However, such long time constants can be very difficult to achieve using pure CMOS circuits~\cite{Qiao_etal17}. 
Memristive devices offer an ideal solution to this limitation. Although such devices are usually treated as non-volatile memories, certain material systems exhibit a rather volatile resistance change after electrical biasing, with temporal scales that can be tuned and matched to biological neural and synaptic dynamics~\cite{Zhang_etal17,Ohno_etal11,Werner_etal16}. Recent demonstrations of volatile memristive devices used to model neural dynamics include the  emulation of nociceptors (i.e., sensory neuron receptors able to detect noxious stimuli)~\cite{Yoon_etal18} and the implementation of spike-timing dependent  learning rules with tunable forgetting rates~\cite{Wang_etal17,Xiong_etal19}. In addition to exploiting the physics of the memristive devices to tune their volatility properties, it is possible to co-design more complex hybrid memristive-CMOS neuromorphic circuits to implement the wide range of time constants needed to model the multiple plasticity phenomena observed in biology (ranging from milliseconds in synaptic short term depression to hours and more in structural plasticity) and crucial for artificial neural processing systems~\cite{Payvand_etal19,Qiao_etal17}.

\paragraph{Don't worry about density.}
Memristive devices are often praised for the small (nano-scale) size, which can be exploited to develop very high density cross-bars~\cite{Pi_etal19} in which the memristive devices are used as a learning synapses~\cite{Xia_etal19}. 
Nevertheless, current high-density approaches are not able to produce learning dynamics sufficiently complex for solving real-world tasks (e.g., with matched temporal scales, or suitable for life-long learning requirements). 
The achievement of such dynamics in a single device requires sophisticated material engineering efforts which are still beyond the current state-of-the-art. 
Conversely, by dismissing the chimera of high density synaptic arrays and co-integrating nano-scale memory elements with mixed signal analog/digital neuromorphic circuits, it is possible to implement sophisticated learning mechanisms that can exploit many features of memristive devices, besides their compact footprint, such as non-volatility, stochasticity, or state-dependent conductance changes. 
Furthermore, combining multiple transistors with one or more memristive devices enables the design of complex synapse circuits that can reduce the effect of variability~\cite{Nair_Indiveri17}, enable the control of stochastic switching behaviors~\cite{Payvand_etal18,Neftci_etal16,Boybat_etal18}, and produce linear or non-linear state-dependent weight-updates~\cite{Brivio_etal19,Diederich_etal18}.

\paragraph{Play it by ear: variability and randomness.}
Memristive devices are affected by both device-to-device and cycle-to-cycle variability~\cite{Fantini_etal13,Suri_Parmar15}. Significant material science and device technology research efforts are being made to minimize such variability~\cite{Schonhals_etal17,Prakash_etal15,Xia_etal19,Fantini_etal13,Govoreanu_etal15,Sheng_etal19}. 
However, rather than fighting these variability effects with different materials or device technologies, neuromorphic systems can be designed to embrace and exploit them~\cite{Payvand_etal19}. 
Examples of theoretical neural processing frameworks that \emph{require} variability can be found in the domain of ensemble learning~\cite{Freund_Schapire97}, reservoir computing~\cite{Jaeger_Haas04} and liquid state machines~\cite{Maass_etal02}. 
Current efforts in neuromorphic engineering for implementing such frameworks to solve spatio-temporal pattern recognition problems rely on the variability provided by transistor device-mismatch effects~\cite{Sheik_etal12a,Richter_etal15,Das_etal18b,Donati_etal18b,Bauer_etal19}. 
Integration of memristive devices with inhomogeneous properties in such architectures can provide a richer set of distributions useful for enhancing the computational abilities of these networks.
Indeed, multiple circuit solutions have already been proposed to better control the shape and parameters of such distributions~\cite{Boybat_etal18,Nair_Indiveri17}.

One important source of variability in the operational parameters of memristive devices is in their switching mechanism. 
In filamentary memristive devices, this mechanism exhibits stochastic behavior which stem from the underlying filament formation process~\cite{Jo_etal10,Gaba_etal13,Ambrogio_etal16a,Yang_etal13}. 
This intrinsic probabilistic property of memristive devices can be exploited for implementing stochastic learning in neuromorphic architectures~\cite{Payvand_etal18,Neftci_etal16,Al-Shedivat_etal15a,Suri_Parmar15,Suri_etal13,Balatti_etal16}, which in turn can be used to implement faithful models of biological cortical microcircuits~\cite{Habenschuss_etal13,Destexhe_Contreras06}, solve memory capacity and classification problems in artificial neural network applications~\cite{Fusi_Senn06,Ginzburg_Sompolinsky94}, and reduce the network sensitivity to their variability~\cite{Payvand_etal18}. 
Recent results on stochastic learning modulated by regularization mechanisms, such as homeostasis or intrinsic plasticity~\cite{Dalgaty_etal19,Yousefzadeh_etal18,Leugering_Pipa18,Neftci_etal16}, present an excellent potential for  exploiting the features memristive devices, even when restricted to binary values.

\paragraph{Don't (hard) limit your devices.}
In the context of deploying always-on learning systems (both artificial and biological) in real-world applications, a critical feature is their memory storage capacity~\cite{Fusi_Abbott07,Ganguli_etal08}. When designing hardware neuromorphic learning system that have practical physical restrictions or limitations on the available resources (such as the number of memory devices integrated in the system, their resolution, precision, or dynamic range) it is important to be aware of the \emph{theoretical} limits that set the bounds of achievable memory capacity and learning performance, independent of the device properties. 

The thorough theoretical analysis on the limits of memory capacity in neural processing systems presented by Fusi and Abbott in 2007~\cite{Fusi_Abbott07} provides essential guiding principles for the construction of artificial learning memristive systems. 
In this analysis, learning models are subdivided into four main categories, according to two key features: the synaptic weight bounds (hard or soft) and the (im)balance of potentiation and depression.
Hard bounds are limits on the synaptic weight values that cannot be exceeded. Soft bounds are limits that can only be reached in the asymptotic limit.  
Typically, in neural network models with hard bounds, the weight update step size is constant and therefore independent of the weight value itself. Conversely soft bounds are introduced by allowing weight updates to depend on synaptic strength and to decrease as they approach the bound itself.

Even though it is clear that in real physical systems hard bounds are unavoidable (e.g., the supply rails in an electronic system), there is evidence that memristive devices exhibit soft bounds~\cite{Frascaroli_etal18}.
Therefore, by combining CMOS circuits with memristive devices, it is possible to design hybrid circuits that can implement and control the devices soft bounds for improving learning at the network level and for improving the overall system performance, e.g., in terms of reduced power consumption and increased memory capacity~\cite{Brivio_etal19}. 
In contrast, it is impossible to precisely balance positive changes of synaptic weights with negative ones in hybrid memristive-CMOS neuromorphic computing systems. Given this unbalanced potentiation and depression property, the longest memory lifetime is achieved thanks to soft bounds, independently of the specific model chosen among those investigated by Fusi and Abbott \cite{Fusi_Abbott07}. 


To best implement the recipe we proposed it is necessary to use the right list of \emph{ingredients}: a combination of memristive devices with multiple complementary features. The recipe \emph{shopping-list} should comprise devices with different properties on retention, endurance, variability, switching currents, on-off ratios, that can be interfaced to analog and digital electronic CMOS circuits. However, even before attempting to bake the final hardware neural processing system, it is important to have access to realistic and faithful device models, so that during the design phase it will be possible to specify the characteristics of both the CMOS and memristive components and understand how to best exploit their processing features for properly modeling the different aspects of plasticity and neural information processing systems.

Once fabricated, these neuromorphic processing systems should implement always-on life-long learning features so that they can adapt to changes in their input signals and keep a proper operating regime. This implies that the hybrid CMOS-memristive neuromorphic system would be updating its synaptic weights continuously, with every learning event. This requires the use of memristive devices that support small gradual conductance changes, and very small currents (e.g., $<1\,\mu$A), to minimize power consumption. In this case, the retention rate of such devices does not need to be extremely long, but should be compatible with the rate of weight update (which can be seen as a ``refresh'' operation) in the system. For example, in typical ``edge'' sensory-processing applications (wearable devices, home automation, surveillance, environmental monitoring, etc.) this could range from milliseconds to seconds or minutes.

On the other hand, once the learning process has terminated or if there is a long pause in the rate of input signals (e.g., during the night in ambient monitoring tasks), then it will be useful to be able to consolidate the memories formed in non-volatile memristive devices with high on-off ratio and long-retention rates. In this case, since this operation would not be as frequent as the weight-update one for the on-line learning case, it would be acceptable to use devices that require larger switching currents, and that have a small number (even two) stable states~\cite{Del-Giudice_etal03}.

To match the time constants of the neural processing system to the dynamics of its input signals, to maintain a stable operating region over long time scales, and to optimize the learning of complex spatio-temporal patterns, it is necessary to implement both fast (short term depression, long term potentiation, long term depression, etc.) and slow (intrinsic, homeostatic, structural) plasticity mechanisms, ``orchestrating'' multiple time-scales in the learning circuits~\cite{Zenke_etal15}. For this it is crucial to be able to use volatile memristive devices that span a wide range of retention rates (e.g., from milliseconds to hours). 

In addition, to increase the memory-capacity of such a system by introducing soft bounds for the synaptic weights, it is necessary to provide a mechanism that can realize the desired state dependence in the synaptic weight-update transfer function~\cite{Brivio_etal19}. This can be achieved by engineering the conductance change properties of the single memristive device, or by designing hybrid memristive-CMOS neuromorphic circuits interfaced with one or more memristive devices~\cite{Bill_Legenstein14,Boybat_etal18}.
Alternatively, one can use multiple binary memristive devices with probabilistic switching in combination with an analog circuit designed to properly control their switching probability.


As evident from the list of ingredients and recipe provided, it is now possible to build ultra low power massively parallel arrays of processing elements that implement ``beyond-von Neumann'', ``in-memory computing'' mixed signal hybrid memristive-CMOS neural processing systems.

It is important to realize that for data-intense processing applications these neuromorphic systems should be used to complement, rather than replace, traditional von Neumann architectures. They could be considered as the \emph{cherry on the cake} of a complex AI inference engine, that enables always-on neural processing, with life-long learning abilities. In this scenario, the hybrid memristive-CMOS neuromorphic computing system would carry out low-power computation acting as alow accuracy predictive ``watch-dog'' to quickly activate more powerful von Neumann architectures for high accuracy recognition, as soon as events of interest are detected.

On the other hand, there are many applications where these hybrid neuromorphic systems would represent both the cherry \emph{and} the cake together: these are IoT, edge-computing, and perception-action tasks that are solved efficiently by biological systems but have been proven to be ``difficult'' for artificial intelligence algorithms~\cite{Plastiras_etal18}. 
This difficulty could be measured with different performance metrics that could range from the physical size and energy consumption requirements to latency, adaptation, and ability to learn in continuous time closed-loop setups. 
By appropriately mixing all the ingredients 
and integrating them with mixed-signal analog/digital neuromorphic systems, it will be possible to produce computing systems that can directly \emph{emulate} their biological counterparts.
This emulation feature, which derives from the exploitation of the physics of the new materials and memory technologies being developed, is the key element for building efficient computing devices that can interact with the environment to solve artificial intelligence tasks in the real physical world, rather than simulating these solutions with general purpose computers. 
In other words, it is not very useful to simulate the bee brain on a supercomputer because it will never fly. 

\begin{acknowledgments}
We wish to acknowledge Melika Payvand and Regina Dittmann for the constructive comments on the manuscript.
The illustration of Fig.~\ref{fig:recipe} was kindly provided by University of Zurich, Information Technology, MELS/SIVIC, Sarah Steinbacher.
This work is supported by the European Research Council (ERC) under the European Union’s Horizon 2020 research and innovation programme grant agreement No 724295 (NeuroAgents).

\end{acknowledgments}

\bibliography{biblio/biblioncs,biblio/bib}

\begin{thebibliography}{80}%
\makeatletter
\providecommand \@ifxundefined [1]{%
 \@ifx{#1\undefined}
}%
\providecommand \@ifnum [1]{%
 \ifnum #1\expandafter \@firstoftwo
 \else \expandafter \@secondoftwo
 \fi
}%
\providecommand \@ifx [1]{%
 \ifx #1\expandafter \@firstoftwo
 \else \expandafter \@secondoftwo
 \fi
}%
\providecommand \natexlab [1]{#1}%
\providecommand \enquote  [1]{``#1''}%
\providecommand \bibnamefont  [1]{#1}%
\providecommand \bibfnamefont [1]{#1}%
\providecommand \citenamefont [1]{#1}%
\providecommand \href@noop [0]{\@secondoftwo}%
\providecommand \href [0]{\begingroup \@sanitize@url \@href}%
\providecommand \@href[1]{\@@startlink{#1}\@@href}%
\providecommand \@@href[1]{\endgroup#1\@@endlink}%
\providecommand \@sanitize@url [0]{\catcode `\\12\catcode `\$12\catcode
  `\&12\catcode `\#12\catcode `\^12\catcode `\_12\catcode `\%12\relax}%
\providecommand \@@startlink[1]{}%
\providecommand \@@endlink[0]{}%
\providecommand \url  [0]{\begingroup\@sanitize@url \@url }%
\providecommand \@url [1]{\endgroup\@href {#1}{\urlprefix }}%
\providecommand \urlprefix  [0]{URL }%
\providecommand \Eprint [0]{\href }%
\providecommand \doibase [0]{http://dx.doi.org/}%
\providecommand \selectlanguage [0]{\@gobble}%
\providecommand \bibinfo  [0]{\@secondoftwo}%
\providecommand \bibfield  [0]{\@secondoftwo}%
\providecommand \translation [1]{[#1]}%
\providecommand \BibitemOpen [0]{}%
\providecommand \bibitemStop [0]{}%
\providecommand \bibitemNoStop [0]{.\EOS\space}%
\providecommand \EOS [0]{\spacefactor3000\relax}%
\providecommand \BibitemShut  [1]{\csname bibitem#1\endcsname}%
\let\auto@bib@innerbib\@empty
\bibitem [{\citenamefont {Neftci}(2018)}]{Neftci18}%
  \BibitemOpen
  \bibfield  {author} {\bibinfo {author} {\bibfnamefont {E.~O.}\ \bibnamefont
  {Neftci}},\ }\bibfield  {title} {\enquote {\bibinfo {title} {Data and power
  efficient intelligence with neuromorphic learning machines},}\ }\href
  {\doibase https://doi.org/10.1016/j.isci.2018.06.010} {\bibfield  {journal}
  {\bibinfo  {journal} {iScience}\ }\textbf {\bibinfo {volume} {5}},\ \bibinfo
  {pages} {52 -- 68} (\bibinfo {year} {2018})}\BibitemShut {NoStop}%
\bibitem [{\citenamefont {Thakur}\ \emph {et~al.}(2018)\citenamefont {Thakur},
  \citenamefont {Molin}, \citenamefont {Cauwenberghs}, \citenamefont
  {Indiveri}, \citenamefont {Kumar}, \citenamefont {Qiao}, \citenamefont
  {Schemmel}, \citenamefont {Wang}, \citenamefont {Chicca}, \citenamefont
  {Olson~Hasler}, \citenamefont {Seo}, \citenamefont {Yu}, \citenamefont {Cao},
  \citenamefont {van Schaik},\ and\ \citenamefont
  {Etienne-Cummings}}]{Thakur_etal18}%
  \BibitemOpen
  \bibfield  {author} {\bibinfo {author} {\bibfnamefont {C.~S.}\ \bibnamefont
  {Thakur}}, \bibinfo {author} {\bibfnamefont {J.~L.}\ \bibnamefont {Molin}},
  \bibinfo {author} {\bibfnamefont {G.}~\bibnamefont {Cauwenberghs}}, \bibinfo
  {author} {\bibfnamefont {G.}~\bibnamefont {Indiveri}}, \bibinfo {author}
  {\bibfnamefont {K.}~\bibnamefont {Kumar}}, \bibinfo {author} {\bibfnamefont
  {N.}~\bibnamefont {Qiao}}, \bibinfo {author} {\bibfnamefont {J.}~\bibnamefont
  {Schemmel}}, \bibinfo {author} {\bibfnamefont {R.}~\bibnamefont {Wang}},
  \bibinfo {author} {\bibfnamefont {E.}~\bibnamefont {Chicca}}, \bibinfo
  {author} {\bibfnamefont {J.}~\bibnamefont {Olson~Hasler}}, \bibinfo {author}
  {\bibfnamefont {J.-s.}\ \bibnamefont {Seo}}, \bibinfo {author} {\bibfnamefont
  {S.}~\bibnamefont {Yu}}, \bibinfo {author} {\bibfnamefont {Y.}~\bibnamefont
  {Cao}}, \bibinfo {author} {\bibfnamefont {A.}~\bibnamefont {van Schaik}}, \
  and\ \bibinfo {author} {\bibfnamefont {R.}~\bibnamefont {Etienne-Cummings}},\
  }\bibfield  {title} {\enquote {\bibinfo {title} {Large-scale neuromorphic
  spiking array processors: A quest to mimic the brain},}\ }\href {\doibase
  10.3389/fnins.2018.00891} {\bibfield  {journal} {\bibinfo  {journal}
  {Frontiers in Neuroscience}\ }\textbf {\bibinfo {volume} {12}},\ \bibinfo
  {pages} {891} (\bibinfo {year} {2018})}\BibitemShut {NoStop}%
\bibitem [{\citenamefont {Li}\ \emph {et~al.}(2018)\citenamefont {Li},
  \citenamefont {Wang}, \citenamefont {Midya}, \citenamefont {Xia},\ and\
  \citenamefont {Yang}}]{Li_etal18c}%
  \BibitemOpen
  \bibfield  {author} {\bibinfo {author} {\bibfnamefont {Y.}~\bibnamefont
  {Li}}, \bibinfo {author} {\bibfnamefont {Z.}~\bibnamefont {Wang}}, \bibinfo
  {author} {\bibfnamefont {R.}~\bibnamefont {Midya}}, \bibinfo {author}
  {\bibfnamefont {Q.}~\bibnamefont {Xia}}, \ and\ \bibinfo {author}
  {\bibfnamefont {J.~J.}\ \bibnamefont {Yang}},\ }\bibfield  {title} {\enquote
  {\bibinfo {title} {Review of memristor devices in neuromorphic computing:
  materials sciences and device challenges},}\ }\href@noop {} {\bibfield
  {journal} {\bibinfo  {journal} {Journal of Physics D: Applied Physics}\
  }\textbf {\bibinfo {volume} {51}},\ \bibinfo {pages} {503002} (\bibinfo
  {year} {2018})}\BibitemShut {NoStop}%
\bibitem [{\citenamefont {van De~Burgt}\ \emph {et~al.}(2018)\citenamefont {van
  De~Burgt}, \citenamefont {Melianas}, \citenamefont {Keene}, \citenamefont
  {Malliaras},\ and\ \citenamefont {Salleo}}]{van-De-Burgt_etal18}%
  \BibitemOpen
  \bibfield  {author} {\bibinfo {author} {\bibfnamefont {Y.}~\bibnamefont {van
  De~Burgt}}, \bibinfo {author} {\bibfnamefont {A.}~\bibnamefont {Melianas}},
  \bibinfo {author} {\bibfnamefont {S.~T.}\ \bibnamefont {Keene}}, \bibinfo
  {author} {\bibfnamefont {G.}~\bibnamefont {Malliaras}}, \ and\ \bibinfo
  {author} {\bibfnamefont {A.}~\bibnamefont {Salleo}},\ }\bibfield  {title}
  {\enquote {\bibinfo {title} {Organic electronics for neuromorphic
  computing},}\ }\href@noop {} {\bibfield  {journal} {\bibinfo  {journal}
  {Nature Electronics}\ }\textbf {\bibinfo {volume} {1}},\ \bibinfo {pages}
  {386--397} (\bibinfo {year} {2018})}\BibitemShut {NoStop}%
\bibitem [{\citenamefont {Burr}\ \emph {et~al.}(2017)\citenamefont {Burr},
  \citenamefont {Shelby}, \citenamefont {Sebastian}, \citenamefont {Kim},
  \citenamefont {Kim}, \citenamefont {Sidler}, \citenamefont {Virwani},
  \citenamefont {Ishii}, \citenamefont {Narayanan}, \citenamefont {Fumarola}
  \emph {et~al.}}]{Burr_etal17}%
  \BibitemOpen
  \bibfield  {author} {\bibinfo {author} {\bibfnamefont {G.~W.}\ \bibnamefont
  {Burr}}, \bibinfo {author} {\bibfnamefont {R.~M.}\ \bibnamefont {Shelby}},
  \bibinfo {author} {\bibfnamefont {A.}~\bibnamefont {Sebastian}}, \bibinfo
  {author} {\bibfnamefont {S.}~\bibnamefont {Kim}}, \bibinfo {author}
  {\bibfnamefont {S.}~\bibnamefont {Kim}}, \bibinfo {author} {\bibfnamefont
  {S.}~\bibnamefont {Sidler}}, \bibinfo {author} {\bibfnamefont
  {K.}~\bibnamefont {Virwani}}, \bibinfo {author} {\bibfnamefont
  {M.}~\bibnamefont {Ishii}}, \bibinfo {author} {\bibfnamefont
  {P.}~\bibnamefont {Narayanan}}, \bibinfo {author} {\bibfnamefont
  {A.}~\bibnamefont {Fumarola}},  \emph {et~al.},\ }\bibfield  {title}
  {\enquote {\bibinfo {title} {Neuromorphic computing using non-volatile
  memory},}\ }\href@noop {} {\bibfield  {journal} {\bibinfo  {journal}
  {Advances in Physics: X}\ }\textbf {\bibinfo {volume} {2}},\ \bibinfo {pages}
  {89--124} (\bibinfo {year} {2017})}\BibitemShut {NoStop}%
\bibitem [{\citenamefont {Mead}(1990)}]{Mead90}%
  \BibitemOpen
  \bibfield  {author} {\bibinfo {author} {\bibfnamefont {C.}~\bibnamefont
  {Mead}},\ }\bibfield  {title} {\enquote {\bibinfo {title} {Neuromorphic
  electronic systems},}\ }\href@noop {} {\bibfield  {journal} {\bibinfo
  {journal} {Proceedings of the {IEEE}}\ }\textbf {\bibinfo {volume} {78}},\
  \bibinfo {pages} {1629--36} (\bibinfo {year} {1990})}\BibitemShut {NoStop}%
\bibitem [{\citenamefont {Furber}\ \emph {et~al.}(2014)\citenamefont {Furber},
  \citenamefont {Galluppi}, \citenamefont {Temple},\ and\ \citenamefont
  {Plana}}]{Furber_etal14}%
  \BibitemOpen
  \bibfield  {author} {\bibinfo {author} {\bibfnamefont {S.}~\bibnamefont
  {Furber}}, \bibinfo {author} {\bibfnamefont {F.}~\bibnamefont {Galluppi}},
  \bibinfo {author} {\bibfnamefont {S.}~\bibnamefont {Temple}}, \ and\ \bibinfo
  {author} {\bibfnamefont {L.}~\bibnamefont {Plana}},\ }\bibfield  {title}
  {\enquote {\bibinfo {title} {The {SpiNNaker} project},}\ }\href {\doibase
  10.1109/JPROC.2014.2304638} {\bibfield  {journal} {\bibinfo  {journal}
  {Proceedings of the IEEE}\ }\textbf {\bibinfo {volume} {102}},\ \bibinfo
  {pages} {652--665} (\bibinfo {year} {2014})}\BibitemShut {NoStop}%
\bibitem [{\citenamefont {Akopyan}\ \emph {et~al.}(2015)\citenamefont
  {Akopyan}, \citenamefont {Sawada}, \citenamefont {Cassidy}, \citenamefont
  {Alvarez-Icaza}, \citenamefont {Arthur}, \citenamefont {Merolla},
  \citenamefont {Imam}, \citenamefont {Nakamura}, \citenamefont {Datta},
  \citenamefont {Nam} \emph {et~al.}}]{Akopyan_etal15}%
  \BibitemOpen
  \bibfield  {author} {\bibinfo {author} {\bibfnamefont {F.}~\bibnamefont
  {Akopyan}}, \bibinfo {author} {\bibfnamefont {J.}~\bibnamefont {Sawada}},
  \bibinfo {author} {\bibfnamefont {A.}~\bibnamefont {Cassidy}}, \bibinfo
  {author} {\bibfnamefont {R.}~\bibnamefont {Alvarez-Icaza}}, \bibinfo {author}
  {\bibfnamefont {J.}~\bibnamefont {Arthur}}, \bibinfo {author} {\bibfnamefont
  {P.}~\bibnamefont {Merolla}}, \bibinfo {author} {\bibfnamefont
  {N.}~\bibnamefont {Imam}}, \bibinfo {author} {\bibfnamefont {Y.}~\bibnamefont
  {Nakamura}}, \bibinfo {author} {\bibfnamefont {P.}~\bibnamefont {Datta}},
  \bibinfo {author} {\bibfnamefont {G.-J.}\ \bibnamefont {Nam}},  \emph
  {et~al.},\ }\bibfield  {title} {\enquote {\bibinfo {title} {Truenorth: Design
  and tool flow of a 65 mw 1 million neuron programmable neurosynaptic chip},}\
  }\href@noop {} {\bibfield  {journal} {\bibinfo  {journal} {IEEE Transactions
  on Computer-Aided Design of Integrated Circuits and Systems}\ }\textbf
  {\bibinfo {volume} {34}},\ \bibinfo {pages} {1537--1557} (\bibinfo {year}
  {2015})}\BibitemShut {NoStop}%
\bibitem [{\citenamefont {Davies}\ \emph {et~al.}(2018)\citenamefont {Davies},
  \citenamefont {Srinivasa}, \citenamefont {Lin}, \citenamefont {Chinya},
  \citenamefont {Cao}, \citenamefont {Choday}, \citenamefont {Dimou},
  \citenamefont {Joshi}, \citenamefont {Imam}, \citenamefont {Jain} \emph
  {et~al.}}]{Davies_etal18}%
  \BibitemOpen
  \bibfield  {author} {\bibinfo {author} {\bibfnamefont {M.}~\bibnamefont
  {Davies}}, \bibinfo {author} {\bibfnamefont {N.}~\bibnamefont {Srinivasa}},
  \bibinfo {author} {\bibfnamefont {T.-H.}\ \bibnamefont {Lin}}, \bibinfo
  {author} {\bibfnamefont {G.}~\bibnamefont {Chinya}}, \bibinfo {author}
  {\bibfnamefont {Y.}~\bibnamefont {Cao}}, \bibinfo {author} {\bibfnamefont
  {S.~H.}\ \bibnamefont {Choday}}, \bibinfo {author} {\bibfnamefont
  {G.}~\bibnamefont {Dimou}}, \bibinfo {author} {\bibfnamefont
  {P.}~\bibnamefont {Joshi}}, \bibinfo {author} {\bibfnamefont
  {N.}~\bibnamefont {Imam}}, \bibinfo {author} {\bibfnamefont {S.}~\bibnamefont
  {Jain}},  \emph {et~al.},\ }\bibfield  {title} {\enquote {\bibinfo {title}
  {Loihi: A neuromorphic manycore processor with on-chip learning},}\
  }\href@noop {} {\bibfield  {journal} {\bibinfo  {journal} {IEEE Micro}\
  }\textbf {\bibinfo {volume} {38}},\ \bibinfo {pages} {82--99} (\bibinfo
  {year} {2018})}\BibitemShut {NoStop}%
\bibitem [{\citenamefont {Ielmini}\ and\ \citenamefont
  {Waser}(2015)}]{Ielmini_Waser15}%
  \BibitemOpen
  \bibfield  {author} {\bibinfo {author} {\bibfnamefont {D.}~\bibnamefont
  {Ielmini}}\ and\ \bibinfo {author} {\bibfnamefont {R.}~\bibnamefont
  {Waser}},\ }\href@noop {} {\emph {\bibinfo {title} {Resistive Switching: From
  Fundamentals of Nanoionic Redox Processes to Memristive Device
  Applications}}}\ (\bibinfo  {publisher} {John Wiley \& Sons},\ \bibinfo
  {year} {2015})\BibitemShut {NoStop}%
\bibitem [{\citenamefont {Boybat}\ \emph {et~al.}(2018)\citenamefont {Boybat},
  \citenamefont {Gallo}, \citenamefont {Moraitis}, \citenamefont {Parnell},
  \citenamefont {Tuma}, \citenamefont {Rajendran}, \citenamefont {Leblebici},
  \citenamefont {Sebastian}, \citenamefont {Eleftheriou} \emph
  {et~al.}}]{Boybat_etal18}%
  \BibitemOpen
  \bibfield  {author} {\bibinfo {author} {\bibfnamefont {I.}~\bibnamefont
  {Boybat}}, \bibinfo {author} {\bibfnamefont {M.~L.}\ \bibnamefont {Gallo}},
  \bibinfo {author} {\bibfnamefont {T.}~\bibnamefont {Moraitis}}, \bibinfo
  {author} {\bibfnamefont {T.}~\bibnamefont {Parnell}}, \bibinfo {author}
  {\bibfnamefont {T.}~\bibnamefont {Tuma}}, \bibinfo {author} {\bibfnamefont
  {B.}~\bibnamefont {Rajendran}}, \bibinfo {author} {\bibfnamefont
  {Y.}~\bibnamefont {Leblebici}}, \bibinfo {author} {\bibfnamefont
  {A.}~\bibnamefont {Sebastian}}, \bibinfo {author} {\bibfnamefont
  {E.}~\bibnamefont {Eleftheriou}},  \emph {et~al.},\ }\bibfield  {title}
  {\enquote {\bibinfo {title} {Neuromorphic computing with multi-memristive
  synapses},}\ }\href@noop {} {\bibfield  {journal} {\bibinfo  {journal}
  {Nature communications}\ }\textbf {\bibinfo {volume} {9}},\ \bibinfo {pages}
  {2514} (\bibinfo {year} {2018})}\BibitemShut {NoStop}%
\bibitem [{\citenamefont {LeCun}, \citenamefont {Bengio},\ and\ \citenamefont
  {Hinton}(2015)}]{LeCun_etal15}%
  \BibitemOpen
  \bibfield  {author} {\bibinfo {author} {\bibfnamefont {Y.}~\bibnamefont
  {LeCun}}, \bibinfo {author} {\bibfnamefont {Y.}~\bibnamefont {Bengio}}, \
  and\ \bibinfo {author} {\bibfnamefont {G.}~\bibnamefont {Hinton}},\
  }\bibfield  {title} {\enquote {\bibinfo {title} {Deep learning},}\
  }\href@noop {} {\bibfield  {journal} {\bibinfo  {journal} {Nature}\ }\textbf
  {\bibinfo {volume} {521}},\ \bibinfo {pages} {436--444} (\bibinfo {year}
  {2015})}\BibitemShut {NoStop}%
\bibitem [{\citenamefont {Schmidhuber}(2015)}]{Schmidhuber15}%
  \BibitemOpen
  \bibfield  {author} {\bibinfo {author} {\bibfnamefont {J.}~\bibnamefont
  {Schmidhuber}},\ }\bibfield  {title} {\enquote {\bibinfo {title} {Deep
  learning in neural networks: An overview},}\ }\href {\doibase
  10.1016/j.neunet.2014.09.003} {\bibfield  {journal} {\bibinfo  {journal}
  {Neural Networks}\ }\textbf {\bibinfo {volume} {61}},\ \bibinfo {pages}
  {85--117} (\bibinfo {year} {2015})}\BibitemShut {NoStop}%
\bibitem [{\citenamefont {Sebastian}, \citenamefont {Gallo},\ and\
  \citenamefont {Eleftheriou}(2019)}]{Sebastian_etal19}%
  \BibitemOpen
  \bibfield  {author} {\bibinfo {author} {\bibfnamefont {A.}~\bibnamefont
  {Sebastian}}, \bibinfo {author} {\bibfnamefont {M.~L.}\ \bibnamefont
  {Gallo}}, \ and\ \bibinfo {author} {\bibfnamefont {E.}~\bibnamefont
  {Eleftheriou}},\ }\bibfield  {title} {\enquote {\bibinfo {title}
  {Computational phase-change memory: beyond von neumann computing},}\ }\href
  {\doibase 10.1088/1361-6463/ab37b6} {\bibfield  {journal} {\bibinfo
  {journal} {Journal of Physics D: Applied Physics}\ }\textbf {\bibinfo
  {volume} {52}},\ \bibinfo {pages} {443002} (\bibinfo {year}
  {2019})}\BibitemShut {NoStop}%
\bibitem [{\citenamefont {Ambrogio}\ \emph {et~al.}(2018)\citenamefont
  {Ambrogio}, \citenamefont {Narayanan}, \citenamefont {Tsai}, \citenamefont
  {Shelby}, \citenamefont {Boybat}, \citenamefont {di~Nolfo}, \citenamefont
  {Sidler}, \citenamefont {Giordano}, \citenamefont {Bodini}, \citenamefont
  {Farinha}, \citenamefont {Killeen}, \citenamefont {Cheng}, \citenamefont
  {Jaoudi},\ and\ \citenamefont {Burr}}]{Ambrogio_etal18}%
  \BibitemOpen
  \bibfield  {author} {\bibinfo {author} {\bibfnamefont {S.}~\bibnamefont
  {Ambrogio}}, \bibinfo {author} {\bibfnamefont {P.}~\bibnamefont {Narayanan}},
  \bibinfo {author} {\bibfnamefont {H.}~\bibnamefont {Tsai}}, \bibinfo {author}
  {\bibfnamefont {R.~M.}\ \bibnamefont {Shelby}}, \bibinfo {author}
  {\bibfnamefont {I.}~\bibnamefont {Boybat}}, \bibinfo {author} {\bibfnamefont
  {C.}~\bibnamefont {di~Nolfo}}, \bibinfo {author} {\bibfnamefont
  {S.}~\bibnamefont {Sidler}}, \bibinfo {author} {\bibfnamefont
  {M.}~\bibnamefont {Giordano}}, \bibinfo {author} {\bibfnamefont
  {M.}~\bibnamefont {Bodini}}, \bibinfo {author} {\bibfnamefont {N.~C.~P.}\
  \bibnamefont {Farinha}}, \bibinfo {author} {\bibfnamefont {B.}~\bibnamefont
  {Killeen}}, \bibinfo {author} {\bibfnamefont {C.}~\bibnamefont {Cheng}},
  \bibinfo {author} {\bibfnamefont {Y.}~\bibnamefont {Jaoudi}}, \ and\ \bibinfo
  {author} {\bibfnamefont {G.~W.}\ \bibnamefont {Burr}},\ }\bibfield  {title}
  {\enquote {\bibinfo {title} {Equivalent-accuracy accelerated neural-network
  training using analogue memory},}\ }\href {\doibase
  10.1038/s41586-018-0180-5} {\bibfield  {journal} {\bibinfo  {journal}
  {Nature}\ }\textbf {\bibinfo {volume} {558}},\ \bibinfo {pages} {60--67}
  (\bibinfo {year} {2018})}\BibitemShut {NoStop}%
\bibitem [{\citenamefont {Dai}\ \emph {et~al.}(2019)\citenamefont {Dai},
  \citenamefont {Zhao}, \citenamefont {Wang}, \citenamefont {Zhang},
  \citenamefont {Fang}, \citenamefont {Jin}, \citenamefont {Shao},\ and\
  \citenamefont {Huang}}]{Dai_etal19}%
  \BibitemOpen
  \bibfield  {author} {\bibinfo {author} {\bibfnamefont {S.}~\bibnamefont
  {Dai}}, \bibinfo {author} {\bibfnamefont {Y.}~\bibnamefont {Zhao}}, \bibinfo
  {author} {\bibfnamefont {Y.}~\bibnamefont {Wang}}, \bibinfo {author}
  {\bibfnamefont {J.}~\bibnamefont {Zhang}}, \bibinfo {author} {\bibfnamefont
  {L.}~\bibnamefont {Fang}}, \bibinfo {author} {\bibfnamefont {S.}~\bibnamefont
  {Jin}}, \bibinfo {author} {\bibfnamefont {Y.}~\bibnamefont {Shao}}, \ and\
  \bibinfo {author} {\bibfnamefont {J.}~\bibnamefont {Huang}},\ }\bibfield
  {title} {\enquote {\bibinfo {title} {Recent advances in transistor-based
  artificial synapses},}\ }\href {\doibase 10.1002/adfm.201903700} {\bibfield
  {journal} {\bibinfo  {journal} {Advanced Functional Materials}\ }\textbf
  {\bibinfo {volume} {0}},\ \bibinfo {pages} {1903700} (\bibinfo {year}
  {2019})},\ \Eprint
  {http://arxiv.org/abs/https://onlinelibrary.wiley.com/doi/pdf/10.1002/adfm.201903700}
  {https://onlinelibrary.wiley.com/doi/pdf/10.1002/adfm.201903700} \BibitemShut
  {NoStop}%
\bibitem [{\citenamefont {Covi}\ \emph {et~al.}(2016)\citenamefont {Covi},
  \citenamefont {Brivio}, \citenamefont {Serb}, \citenamefont {Prodromakis},
  \citenamefont {Fanciulli},\ and\ \citenamefont {Spiga}}]{Covi_etal16}%
  \BibitemOpen
  \bibfield  {author} {\bibinfo {author} {\bibfnamefont {E.}~\bibnamefont
  {Covi}}, \bibinfo {author} {\bibfnamefont {S.}~\bibnamefont {Brivio}},
  \bibinfo {author} {\bibfnamefont {A.}~\bibnamefont {Serb}}, \bibinfo {author}
  {\bibfnamefont {T.}~\bibnamefont {Prodromakis}}, \bibinfo {author}
  {\bibfnamefont {M.}~\bibnamefont {Fanciulli}}, \ and\ \bibinfo {author}
  {\bibfnamefont {S.}~\bibnamefont {Spiga}},\ }\bibfield  {title} {\enquote
  {\bibinfo {title} {Analog memristive synapse in spiking networks implementing
  unsupervised learning},}\ }\href@noop {} {\bibfield  {journal} {\bibinfo
  {journal} {Frontiers in neuroscience}\ }\textbf {\bibinfo {volume} {10}},\
  \bibinfo {pages} {1--13} (\bibinfo {year} {2016})}\BibitemShut {NoStop}%
\bibitem [{\citenamefont {Jo}\ \emph {et~al.}(2010)\citenamefont {Jo},
  \citenamefont {Chang}, \citenamefont {Ebong}, \citenamefont {Bhadviya},
  \citenamefont {Mazumder},\ and\ \citenamefont {Lu}}]{Jo_etal10}%
  \BibitemOpen
  \bibfield  {author} {\bibinfo {author} {\bibfnamefont {S.~H.}\ \bibnamefont
  {Jo}}, \bibinfo {author} {\bibfnamefont {T.}~\bibnamefont {Chang}}, \bibinfo
  {author} {\bibfnamefont {I.}~\bibnamefont {Ebong}}, \bibinfo {author}
  {\bibfnamefont {B.~B.}\ \bibnamefont {Bhadviya}}, \bibinfo {author}
  {\bibfnamefont {P.}~\bibnamefont {Mazumder}}, \ and\ \bibinfo {author}
  {\bibfnamefont {W.}~\bibnamefont {Lu}},\ }\bibfield  {title} {\enquote
  {\bibinfo {title} {Nanoscale memristor device as synapse in neuromorphic
  systems},}\ }\href@noop {} {\bibfield  {journal} {\bibinfo  {journal} {Nano
  letters}\ }\textbf {\bibinfo {volume} {10}},\ \bibinfo {pages} {1297--1301}
  (\bibinfo {year} {2010})}\BibitemShut {NoStop}%
\bibitem [{\citenamefont {Yang}\ and\ \citenamefont {Xia}(2017)}]{Yang_etal17}%
  \BibitemOpen
  \bibfield  {author} {\bibinfo {author} {\bibfnamefont {J.~J.}\ \bibnamefont
  {Yang}}\ and\ \bibinfo {author} {\bibfnamefont {Q.}~\bibnamefont {Xia}},\
  }\bibfield  {title} {\enquote {\bibinfo {title} {Organic electronics:
  Battery-like artificial synapses},}\ }\href@noop {} {\bibfield  {journal}
  {\bibinfo  {journal} {Nature materials}\ }\textbf {\bibinfo {volume} {16}},\
  \bibinfo {pages} {396} (\bibinfo {year} {2017})}\BibitemShut {NoStop}%
\bibitem [{\citenamefont {Berdan}\ \emph {et~al.}(2016)\citenamefont {Berdan},
  \citenamefont {Vasilaki}, \citenamefont {Khiat}, \citenamefont {Indiveri},
  \citenamefont {Serb},\ and\ \citenamefont {Prodromakis}}]{Berdan_etal16}%
  \BibitemOpen
  \bibfield  {author} {\bibinfo {author} {\bibfnamefont {R.}~\bibnamefont
  {Berdan}}, \bibinfo {author} {\bibfnamefont {E.}~\bibnamefont {Vasilaki}},
  \bibinfo {author} {\bibfnamefont {A.}~\bibnamefont {Khiat}}, \bibinfo
  {author} {\bibfnamefont {G.}~\bibnamefont {Indiveri}}, \bibinfo {author}
  {\bibfnamefont {A.}~\bibnamefont {Serb}}, \ and\ \bibinfo {author}
  {\bibfnamefont {T.}~\bibnamefont {Prodromakis}},\ }\bibfield  {title}
  {\enquote {\bibinfo {title} {Emulating short-term synaptic dynamics with
  memristive devices},}\ }\href {\doibase 10.1038/srep18639} {\bibfield
  {journal} {\bibinfo  {journal} {Scientific Reports}\ }\textbf {\bibinfo
  {volume} {6}},\ \bibinfo {pages} {1--9} (\bibinfo {year} {2016})}\BibitemShut
  {NoStop}%
\bibitem [{\citenamefont {Chicca}\ \emph {et~al.}(2014)\citenamefont {Chicca},
  \citenamefont {Stefanini}, \citenamefont {Bartolozzi},\ and\ \citenamefont
  {Indiveri}}]{Chicca_etal14}%
  \BibitemOpen
  \bibfield  {author} {\bibinfo {author} {\bibfnamefont {E.}~\bibnamefont
  {Chicca}}, \bibinfo {author} {\bibfnamefont {F.}~\bibnamefont {Stefanini}},
  \bibinfo {author} {\bibfnamefont {C.}~\bibnamefont {Bartolozzi}}, \ and\
  \bibinfo {author} {\bibfnamefont {G.}~\bibnamefont {Indiveri}},\ }\bibfield
  {title} {\enquote {\bibinfo {title} {Neuromorphic electronic circuits for
  building autonomous cognitive systems},}\ }\href@noop {} {\bibfield
  {journal} {\bibinfo  {journal} {Proceedings of the {IEEE}}\ }\textbf
  {\bibinfo {volume} {102}},\ \bibinfo {pages} {1367--1388} (\bibinfo {year}
  {2014})}\BibitemShut {NoStop}%
\bibitem [{\citenamefont {Indiveri}\ and\ \citenamefont
  {Liu}(2015)}]{Indiveri_Liu15}%
  \BibitemOpen
  \bibfield  {author} {\bibinfo {author} {\bibfnamefont {G.}~\bibnamefont
  {Indiveri}}\ and\ \bibinfo {author} {\bibfnamefont {S.-C.}\ \bibnamefont
  {Liu}},\ }\bibfield  {title} {\enquote {\bibinfo {title} {Memory and
  information processing in neuromorphic systems},}\ }\href {\doibase
  10.1109/JPROC.2015.2444094} {\bibfield  {journal} {\bibinfo  {journal}
  {Proceedings of the {IEEE}}\ }\textbf {\bibinfo {volume} {103}},\ \bibinfo
  {pages} {1379--1397} (\bibinfo {year} {2015})}\BibitemShut {NoStop}%
\bibitem [{\citenamefont {Indiveri}\ and\ \citenamefont
  {Sandamirskaya}(2019)}]{Indiveri_Sandamirskaya19}%
  \BibitemOpen
  \bibfield  {author} {\bibinfo {author} {\bibfnamefont {G.}~\bibnamefont
  {Indiveri}}\ and\ \bibinfo {author} {\bibfnamefont {Y.}~\bibnamefont
  {Sandamirskaya}},\ }\bibfield  {title} {\enquote {\bibinfo {title} {The
  importance of space and time for signal processing in neuromorphic agents},}\
  }\href {\doibase 10.1109/MSP.2019.2928376} {\bibfield  {journal} {\bibinfo
  {journal} {{IEEE} Signal Processing Magazine}\ }\textbf {\bibinfo {volume}
  {36}},\ \bibinfo {pages} {16--28} (\bibinfo {year} {2019})}\BibitemShut
  {NoStop}%
\bibitem [{\citenamefont {Merolla}\ \emph {et~al.}(2014)\citenamefont
  {Merolla}, \citenamefont {Arthur}, \citenamefont {Alvarez-Icaza},
  \citenamefont {Cassidy}, \citenamefont {Sawada}, \citenamefont {Akopyan},
  \citenamefont {Jackson}, \citenamefont {Imam}, \citenamefont {Guo},
  \citenamefont {Nakamura}, \citenamefont {Brezzo}, \citenamefont {Vo},
  \citenamefont {Esser}, \citenamefont {Appuswamy}, \citenamefont {Taba},
  \citenamefont {Amir}, \citenamefont {Flickner}, \citenamefont {Risk},
  \citenamefont {Manohar},\ and\ \citenamefont {Modha}}]{Merolla_etal14a}%
  \BibitemOpen
  \bibfield  {author} {\bibinfo {author} {\bibfnamefont {P.~A.}\ \bibnamefont
  {Merolla}}, \bibinfo {author} {\bibfnamefont {J.~V.}\ \bibnamefont {Arthur}},
  \bibinfo {author} {\bibfnamefont {R.}~\bibnamefont {Alvarez-Icaza}}, \bibinfo
  {author} {\bibfnamefont {A.~S.}\ \bibnamefont {Cassidy}}, \bibinfo {author}
  {\bibfnamefont {J.}~\bibnamefont {Sawada}}, \bibinfo {author} {\bibfnamefont
  {F.}~\bibnamefont {Akopyan}}, \bibinfo {author} {\bibfnamefont {B.~L.}\
  \bibnamefont {Jackson}}, \bibinfo {author} {\bibfnamefont {N.}~\bibnamefont
  {Imam}}, \bibinfo {author} {\bibfnamefont {C.}~\bibnamefont {Guo}}, \bibinfo
  {author} {\bibfnamefont {Y.}~\bibnamefont {Nakamura}}, \bibinfo {author}
  {\bibfnamefont {B.}~\bibnamefont {Brezzo}}, \bibinfo {author} {\bibfnamefont
  {I.}~\bibnamefont {Vo}}, \bibinfo {author} {\bibfnamefont {S.~K.}\
  \bibnamefont {Esser}}, \bibinfo {author} {\bibfnamefont {R.}~\bibnamefont
  {Appuswamy}}, \bibinfo {author} {\bibfnamefont {B.}~\bibnamefont {Taba}},
  \bibinfo {author} {\bibfnamefont {A.}~\bibnamefont {Amir}}, \bibinfo {author}
  {\bibfnamefont {M.~D.}\ \bibnamefont {Flickner}}, \bibinfo {author}
  {\bibfnamefont {W.~P.}\ \bibnamefont {Risk}}, \bibinfo {author}
  {\bibfnamefont {R.}~\bibnamefont {Manohar}}, \ and\ \bibinfo {author}
  {\bibfnamefont {D.~S.}\ \bibnamefont {Modha}},\ }\bibfield  {title} {\enquote
  {\bibinfo {title} {A million spiking-neuron integrated circuit with a
  scalable communication network and interface},}\ }\href@noop {} {\bibfield
  {journal} {\bibinfo  {journal} {Science}\ }\textbf {\bibinfo {volume}
  {345}},\ \bibinfo {pages} {668--673} (\bibinfo {year} {2014})}\BibitemShut
  {NoStop}%
\bibitem [{\citenamefont {Backus}(1978)}]{Backus78}%
  \BibitemOpen
  \bibfield  {author} {\bibinfo {author} {\bibfnamefont {J.}~\bibnamefont
  {Backus}},\ }\bibfield  {title} {\enquote {\bibinfo {title} {Can programming
  be liberated from the von neumann style?: a functional style and its algebra
  of programs},}\ }\href {\doibase 10.1145/359576.359579} {\bibfield  {journal}
  {\bibinfo  {journal} {Communications of the ACM}\ }\textbf {\bibinfo {volume}
  {21}},\ \bibinfo {pages} {613--641} (\bibinfo {year} {1978})}\BibitemShut
  {NoStop}%
\bibitem [{\citenamefont {Indiveri}\ \emph {et~al.}(2013)\citenamefont
  {Indiveri}, \citenamefont {Linares-Barranco}, \citenamefont {Legenstein},
  \citenamefont {Deligeorgis},\ and\ \citenamefont
  {Prodromakis}}]{Indiveri_etal13}%
  \BibitemOpen
  \bibfield  {author} {\bibinfo {author} {\bibfnamefont {G.}~\bibnamefont
  {Indiveri}}, \bibinfo {author} {\bibfnamefont {B.}~\bibnamefont
  {Linares-Barranco}}, \bibinfo {author} {\bibfnamefont {R.}~\bibnamefont
  {Legenstein}}, \bibinfo {author} {\bibfnamefont {G.}~\bibnamefont
  {Deligeorgis}}, \ and\ \bibinfo {author} {\bibfnamefont {T.}~\bibnamefont
  {Prodromakis}},\ }\bibfield  {title} {\enquote {\bibinfo {title} {Integration
  of nanoscale memristor synapses in neuromorphic computing architectures},}\
  }\href {\doibase 10.1088/0957-4484/24/38/384010} {\bibfield  {journal}
  {\bibinfo  {journal} {Nanotechnology}\ }\textbf {\bibinfo {volume} {24}},\
  \bibinfo {pages} {384010} (\bibinfo {year} {2013})}\BibitemShut {NoStop}%
\bibitem [{\citenamefont {Von~Neumann}(1993)}]{Von-Neumann93}%
  \BibitemOpen
  \bibfield  {author} {\bibinfo {author} {\bibfnamefont {J.}~\bibnamefont
  {Von~Neumann}},\ }\bibfield  {title} {\enquote {\bibinfo {title} {First draft
  of a report on the edvac},}\ }\href@noop {} {\bibfield  {journal} {\bibinfo
  {journal} {IEEE Annals of the History of Computing}\ }\textbf {\bibinfo
  {volume} {15}},\ \bibinfo {pages} {27--75} (\bibinfo {year}
  {1993})}\BibitemShut {NoStop}%
\bibitem [{\citenamefont {McCulloch}\ and\ \citenamefont
  {Pitts}(1943)}]{McCulloch_Pitts43}%
  \BibitemOpen
  \bibfield  {author} {\bibinfo {author} {\bibfnamefont {W.}~\bibnamefont
  {McCulloch}}\ and\ \bibinfo {author} {\bibfnamefont {W.}~\bibnamefont
  {Pitts}},\ }\bibfield  {title} {\enquote {\bibinfo {title} {A logical
  calculus of the ideas immanent in nervous activity},}\ }\href@noop {}
  {\bibfield  {journal} {\bibinfo  {journal} {Bull. Math. Biophys.}\ }\textbf
  {\bibinfo {volume} {5}},\ \bibinfo {pages} {115--133} (\bibinfo {year}
  {1943})}\BibitemShut {NoStop}%
\bibitem [{\citenamefont {Shadlen}\ and\ \citenamefont
  {Movshon}(1999)}]{Shadlen_Movshon99}%
  \BibitemOpen
  \bibfield  {author} {\bibinfo {author} {\bibfnamefont {M.}~\bibnamefont
  {Shadlen}}\ and\ \bibinfo {author} {\bibfnamefont {J.}~\bibnamefont
  {Movshon}},\ }\bibfield  {title} {\enquote {\bibinfo {title} {Synchrony
  unbound: a critical evaluation of the temporal binding hypothesis},}\
  }\href@noop {} {\bibfield  {journal} {\bibinfo  {journal} {Neuron}\ }\textbf
  {\bibinfo {volume} {24}},\ \bibinfo {pages} {67--77} (\bibinfo {year}
  {1999})}\BibitemShut {NoStop}%
\bibitem [{\citenamefont {Boahen}(1998)}]{Boahen98}%
  \BibitemOpen
  \bibfield  {author} {\bibinfo {author} {\bibfnamefont {K.}~\bibnamefont
  {Boahen}},\ }\bibfield  {title} {\enquote {\bibinfo {title} {Communicating
  neuronal ensembles between neuromorphic chips},}\ }in\ \href@noop {} {\emph
  {\bibinfo {booktitle} {Neuromorphic Systems Engineering}}},\ \bibinfo
  {editor} {edited by\ \bibinfo {editor} {\bibfnamefont {T.}~\bibnamefont
  {Lande}}}\ (\bibinfo  {publisher} {Kluwer Academic},\ \bibinfo {address}
  {Norwell, MA},\ \bibinfo {year} {1998})\ pp.\ \bibinfo {pages}
  {229--259}\BibitemShut {NoStop}%
\bibitem [{\citenamefont {Moradi}\ \emph {et~al.}(2018)\citenamefont {Moradi},
  \citenamefont {Qiao}, \citenamefont {Stefanini},\ and\ \citenamefont
  {Indiveri}}]{Moradi_etal18}%
  \BibitemOpen
  \bibfield  {author} {\bibinfo {author} {\bibfnamefont {S.}~\bibnamefont
  {Moradi}}, \bibinfo {author} {\bibfnamefont {N.}~\bibnamefont {Qiao}},
  \bibinfo {author} {\bibfnamefont {F.}~\bibnamefont {Stefanini}}, \ and\
  \bibinfo {author} {\bibfnamefont {G.}~\bibnamefont {Indiveri}},\ }\bibfield
  {title} {\enquote {\bibinfo {title} {A scalable multicore architecture with
  heterogeneous memory structures for dynamic neuromorphic asynchronous
  processors ({DYNAPs})},}\ }\href@noop {} {\bibfield  {journal} {\bibinfo
  {journal} {Biomedical Circuits and Systems, {IEEE} Transactions on}\ }\textbf
  {\bibinfo {volume} {12}},\ \bibinfo {pages} {106--122} (\bibinfo {year}
  {2018})}\BibitemShut {NoStop}%
\bibitem [{\citenamefont {Qiao}, \citenamefont {Bartolozzi},\ and\
  \citenamefont {Indiveri}(2017)}]{Qiao_etal17}%
  \BibitemOpen
  \bibfield  {author} {\bibinfo {author} {\bibfnamefont {N.}~\bibnamefont
  {Qiao}}, \bibinfo {author} {\bibfnamefont {C.}~\bibnamefont {Bartolozzi}}, \
  and\ \bibinfo {author} {\bibfnamefont {G.}~\bibnamefont {Indiveri}},\
  }\bibfield  {title} {\enquote {\bibinfo {title} {An ultralow leakage synaptic
  scaling homeostatic plasticity circuit with configurable time scales up to
  100 ks},}\ }\href {\doibase 10.1109/TBCAS.2017.2754383} {\bibfield  {journal}
  {\bibinfo  {journal} {{IEEE} Transactions on Biomedical Circuits and
  Systems}\ } (\bibinfo {year} {2017}),\
  10.1109/TBCAS.2017.2754383}\BibitemShut {NoStop}%
\bibitem [{\citenamefont {Zhang}\ \emph {et~al.}(2017)\citenamefont {Zhang},
  \citenamefont {Liu}, \citenamefont {Zhao}, \citenamefont {Wu}, \citenamefont
  {Wu}, \citenamefont {Wang}, \citenamefont {Cao}, \citenamefont {Fang},
  \citenamefont {Lv}, \citenamefont {Long}, \citenamefont {Liu},\ and\
  \citenamefont {Liu}}]{Zhang_etal17}%
  \BibitemOpen
  \bibfield  {author} {\bibinfo {author} {\bibfnamefont {X.}~\bibnamefont
  {Zhang}}, \bibinfo {author} {\bibfnamefont {S.}~\bibnamefont {Liu}}, \bibinfo
  {author} {\bibfnamefont {X.}~\bibnamefont {Zhao}}, \bibinfo {author}
  {\bibfnamefont {F.}~\bibnamefont {Wu}}, \bibinfo {author} {\bibfnamefont
  {Q.}~\bibnamefont {Wu}}, \bibinfo {author} {\bibfnamefont {W.}~\bibnamefont
  {Wang}}, \bibinfo {author} {\bibfnamefont {R.}~\bibnamefont {Cao}}, \bibinfo
  {author} {\bibfnamefont {Y.}~\bibnamefont {Fang}}, \bibinfo {author}
  {\bibfnamefont {H.}~\bibnamefont {Lv}}, \bibinfo {author} {\bibfnamefont
  {S.}~\bibnamefont {Long}}, \bibinfo {author} {\bibfnamefont {Q.}~\bibnamefont
  {Liu}}, \ and\ \bibinfo {author} {\bibfnamefont {M.}~\bibnamefont {Liu}},\
  }\bibfield  {title} {\enquote {\bibinfo {title} {Emulating short-term and
  long-term plasticity of bio-synapse based on cu/a-si/pt memristor},}\
  }\href@noop {} {\bibfield  {journal} {\bibinfo  {journal} {{IEEE} Electron
  Device Letters}\ }\textbf {\bibinfo {volume} {38}},\ \bibinfo {pages}
  {1208--1211} (\bibinfo {year} {2017})}\BibitemShut {NoStop}%
\bibitem [{\citenamefont {Ohno}\ \emph {et~al.}(2011)\citenamefont {Ohno},
  \citenamefont {Hasegawa}, \citenamefont {Tsuruoka}, \citenamefont {Terabe},
  \citenamefont {Gimzewski},\ and\ \citenamefont {Aono}}]{Ohno_etal11}%
  \BibitemOpen
  \bibfield  {author} {\bibinfo {author} {\bibfnamefont {T.}~\bibnamefont
  {Ohno}}, \bibinfo {author} {\bibfnamefont {T.}~\bibnamefont {Hasegawa}},
  \bibinfo {author} {\bibfnamefont {T.}~\bibnamefont {Tsuruoka}}, \bibinfo
  {author} {\bibfnamefont {K.}~\bibnamefont {Terabe}}, \bibinfo {author}
  {\bibfnamefont {J.}~\bibnamefont {Gimzewski}}, \ and\ \bibinfo {author}
  {\bibfnamefont {M.}~\bibnamefont {Aono}},\ }\bibfield  {title} {\enquote
  {\bibinfo {title} {Short-term plasticity and long-term potentiation mimicked
  in single inorganic synapses},}\ }\href@noop {} {\bibfield  {journal}
  {\bibinfo  {journal} {Nature Materials}\ }\textbf {\bibinfo {volume} {10}},\
  \bibinfo {pages} {591--595} (\bibinfo {year} {2011})}\BibitemShut {NoStop}%
\bibitem [{\citenamefont {Werner}\ \emph {et~al.}(2016)\citenamefont {Werner},
  \citenamefont {Vianello}, \citenamefont {Bichler}, \citenamefont {Grossi},
  \citenamefont {Nowak}, \citenamefont {Nodin}, \citenamefont {Yvert},
  \citenamefont {Salvo},\ and\ \citenamefont {Perniola}}]{Werner_etal16}%
  \BibitemOpen
  \bibfield  {author} {\bibinfo {author} {\bibfnamefont {T.}~\bibnamefont
  {Werner}}, \bibinfo {author} {\bibfnamefont {E.}~\bibnamefont {Vianello}},
  \bibinfo {author} {\bibfnamefont {O.}~\bibnamefont {Bichler}}, \bibinfo
  {author} {\bibfnamefont {A.}~\bibnamefont {Grossi}}, \bibinfo {author}
  {\bibfnamefont {E.}~\bibnamefont {Nowak}}, \bibinfo {author} {\bibfnamefont
  {J.-F.}\ \bibnamefont {Nodin}}, \bibinfo {author} {\bibfnamefont
  {B.}~\bibnamefont {Yvert}}, \bibinfo {author} {\bibfnamefont {B.~D.}\
  \bibnamefont {Salvo}}, \ and\ \bibinfo {author} {\bibfnamefont
  {L.}~\bibnamefont {Perniola}},\ }\bibfield  {title} {\enquote {\bibinfo
  {title} {Experimental demonstration of short and long term synaptic
  plasticity using ox{RAM} multi k-bit arrays for reliable detection in highly
  noisy input data},}\ }in\ \href@noop {} {\emph {\bibinfo {booktitle} {2016
  {IEEE} International Electron Devices Meeting ({IEDM})}}}\ (\bibinfo
  {organization} {{IEEE}},\ \bibinfo {year} {2016})\ pp.\ \bibinfo {pages}
  {16--6}\BibitemShut {NoStop}%
\bibitem [{\citenamefont {Yoon}\ \emph {et~al.}(2018)\citenamefont {Yoon},
  \citenamefont {Jung}, \citenamefont {Wang}, \citenamefont {Kim},
  \citenamefont {Wu}, \citenamefont {Ravichandran}, \citenamefont {Xia},
  \citenamefont {Hwang},\ and\ \citenamefont {Yang}}]{Yoon_etal18}%
  \BibitemOpen
  \bibfield  {author} {\bibinfo {author} {\bibfnamefont {J.}~\bibnamefont
  {Yoon}}, \bibinfo {author} {\bibfnamefont {H.}~\bibnamefont {Jung}}, \bibinfo
  {author} {\bibfnamefont {Z.}~\bibnamefont {Wang}}, \bibinfo {author}
  {\bibfnamefont {K.~M.}\ \bibnamefont {Kim}}, \bibinfo {author} {\bibfnamefont
  {H.}~\bibnamefont {Wu}}, \bibinfo {author} {\bibfnamefont {V.}~\bibnamefont
  {Ravichandran}}, \bibinfo {author} {\bibfnamefont {Q.}~\bibnamefont {Xia}},
  \bibinfo {author} {\bibfnamefont {C.~S.}\ \bibnamefont {Hwang}}, \ and\
  \bibinfo {author} {\bibfnamefont {J.~J.}\ \bibnamefont {Yang}},\ }\bibfield
  {title} {\enquote {\bibinfo {title} {An artificial nociceptor based on a
  diffusive memristor},}\ }\href@noop {} {\bibfield  {journal} {\bibinfo
  {journal} {Nature communications}\ }\textbf {\bibinfo {volume} {9}},\
  \bibinfo {pages} {417} (\bibinfo {year} {2018})}\BibitemShut {NoStop}%
\bibitem [{\citenamefont {Wang}\ \emph {et~al.}(2017)\citenamefont {Wang},
  \citenamefont {Joshi}, \citenamefont {Savel’ev}, \citenamefont {Jiang},
  \citenamefont {Midya}, \citenamefont {Lin}, \citenamefont {Hu}, \citenamefont
  {Ge}, \citenamefont {Strachan}, \citenamefont {Li}, \citenamefont {Wu},
  \citenamefont {Barnell}, \citenamefont {Li}, \citenamefont {Xin},
  \citenamefont {Williams}, \citenamefont {Xia},\ and\ \citenamefont
  {Yang}}]{Wang_etal17}%
  \BibitemOpen
  \bibfield  {author} {\bibinfo {author} {\bibfnamefont {Z.}~\bibnamefont
  {Wang}}, \bibinfo {author} {\bibfnamefont {S.}~\bibnamefont {Joshi}},
  \bibinfo {author} {\bibfnamefont {S.~E.}\ \bibnamefont {Savel’ev}},
  \bibinfo {author} {\bibfnamefont {H.}~\bibnamefont {Jiang}}, \bibinfo
  {author} {\bibfnamefont {R.}~\bibnamefont {Midya}}, \bibinfo {author}
  {\bibfnamefont {P.}~\bibnamefont {Lin}}, \bibinfo {author} {\bibfnamefont
  {M.}~\bibnamefont {Hu}}, \bibinfo {author} {\bibfnamefont {N.}~\bibnamefont
  {Ge}}, \bibinfo {author} {\bibfnamefont {J.~P.}\ \bibnamefont {Strachan}},
  \bibinfo {author} {\bibfnamefont {Z.}~\bibnamefont {Li}}, \bibinfo {author}
  {\bibfnamefont {Q.}~\bibnamefont {Wu}}, \bibinfo {author} {\bibfnamefont
  {M.}~\bibnamefont {Barnell}}, \bibinfo {author} {\bibfnamefont {G.-L.}\
  \bibnamefont {Li}}, \bibinfo {author} {\bibfnamefont {H.~L.}\ \bibnamefont
  {Xin}}, \bibinfo {author} {\bibfnamefont {R.~S.}\ \bibnamefont {Williams}},
  \bibinfo {author} {\bibfnamefont {Q.}~\bibnamefont {Xia}}, \ and\ \bibinfo
  {author} {\bibfnamefont {J.~J.}\ \bibnamefont {Yang}},\ }\bibfield  {title}
  {\enquote {\bibinfo {title} {Memristors with diffusive dynamics as synaptic
  emulators for neuromorphic computing},}\ }\href@noop {} {\bibfield  {journal}
  {\bibinfo  {journal} {Nature materials}\ }\textbf {\bibinfo {volume} {16}},\
  \bibinfo {pages} {101} (\bibinfo {year} {2017})}\BibitemShut {NoStop}%
\bibitem [{\citenamefont {Xiong}\ \emph {et~al.}(2019)\citenamefont {Xiong},
  \citenamefont {Yang}, \citenamefont {Shaibo}, \citenamefont {Huang},
  \citenamefont {He}, \citenamefont {Zhou},\ and\ \citenamefont
  {Guo}}]{Xiong_etal19}%
  \BibitemOpen
  \bibfield  {author} {\bibinfo {author} {\bibfnamefont {J.}~\bibnamefont
  {Xiong}}, \bibinfo {author} {\bibfnamefont {R.}~\bibnamefont {Yang}},
  \bibinfo {author} {\bibfnamefont {J.}~\bibnamefont {Shaibo}}, \bibinfo
  {author} {\bibfnamefont {H.~M.}\ \bibnamefont {Huang}}, \bibinfo {author}
  {\bibfnamefont {H.~K.}\ \bibnamefont {He}}, \bibinfo {author} {\bibfnamefont
  {W.}~\bibnamefont {Zhou}}, \ and\ \bibinfo {author} {\bibfnamefont
  {X.}~\bibnamefont {Guo}},\ }\bibfield  {title} {\enquote {\bibinfo {title}
  {Bienenstock, cooper, and munro learning rules realized in second-order
  memristors with tunable forgetting rate},}\ }\href@noop {} {\bibfield
  {journal} {\bibinfo  {journal} {Advanced Functional Materials}\ }\textbf
  {\bibinfo {volume} {29}},\ \bibinfo {pages} {1807316} (\bibinfo {year}
  {2019})}\BibitemShut {NoStop}%
\bibitem [{\citenamefont {Payvand}\ \emph {et~al.}(2019)\citenamefont
  {Payvand}, \citenamefont {Nair}, \citenamefont {M{\"u}ller},\ and\
  \citenamefont {Indiveri}}]{Payvand_etal19}%
  \BibitemOpen
  \bibfield  {author} {\bibinfo {author} {\bibfnamefont {M.}~\bibnamefont
  {Payvand}}, \bibinfo {author} {\bibfnamefont {M.~V.}\ \bibnamefont {Nair}},
  \bibinfo {author} {\bibfnamefont {L.~K.}\ \bibnamefont {M{\"u}ller}}, \ and\
  \bibinfo {author} {\bibfnamefont {G.}~\bibnamefont {Indiveri}},\ }\bibfield
  {title} {\enquote {\bibinfo {title} {A neuromorphic systems approach to
  in-memory computing with non-ideal memristive devices: From mitigation to
  exploitation},}\ }\href@noop {} {\bibfield  {journal} {\bibinfo  {journal}
  {Faraday Discussions}\ }\textbf {\bibinfo {volume} {213}},\ \bibinfo {pages}
  {487--510} (\bibinfo {year} {2019})}\BibitemShut {NoStop}%
\bibitem [{\citenamefont {Pi}\ \emph {et~al.}(2019)\citenamefont {Pi},
  \citenamefont {Li}, \citenamefont {Jiang}, \citenamefont {Xia}, \citenamefont
  {Xin}, \citenamefont {Yang},\ and\ \citenamefont {Xia}}]{Pi_etal19}%
  \BibitemOpen
  \bibfield  {author} {\bibinfo {author} {\bibfnamefont {S.}~\bibnamefont
  {Pi}}, \bibinfo {author} {\bibfnamefont {C.}~\bibnamefont {Li}}, \bibinfo
  {author} {\bibfnamefont {H.}~\bibnamefont {Jiang}}, \bibinfo {author}
  {\bibfnamefont {W.}~\bibnamefont {Xia}}, \bibinfo {author} {\bibfnamefont
  {H.}~\bibnamefont {Xin}}, \bibinfo {author} {\bibfnamefont {J.~J.}\
  \bibnamefont {Yang}}, \ and\ \bibinfo {author} {\bibfnamefont
  {Q.}~\bibnamefont {Xia}},\ }\bibfield  {title} {\enquote {\bibinfo {title}
  {Memristor crossbar arrays with 6-nm half-pitch and 2-nm critical
  dimension},}\ }\href@noop {} {\bibfield  {journal} {\bibinfo  {journal}
  {Nature nanotechnology}\ }\textbf {\bibinfo {volume} {14}},\ \bibinfo {pages}
  {35} (\bibinfo {year} {2019})}\BibitemShut {NoStop}%
\bibitem [{\citenamefont {Xia}\ and\ \citenamefont {Yang}(2019)}]{Xia_etal19}%
  \BibitemOpen
  \bibfield  {author} {\bibinfo {author} {\bibfnamefont {Q.}~\bibnamefont
  {Xia}}\ and\ \bibinfo {author} {\bibfnamefont {J.~J.}\ \bibnamefont {Yang}},\
  }\bibfield  {title} {\enquote {\bibinfo {title} {Memristive crossbar arrays
  for brain-inspired computing},}\ }\href@noop {} {\bibfield  {journal}
  {\bibinfo  {journal} {Nature materials}\ }\textbf {\bibinfo {volume} {18}},\
  \bibinfo {pages} {309} (\bibinfo {year} {2019})}\BibitemShut {NoStop}%
\bibitem [{\citenamefont {Nair}\ and\ \citenamefont
  {Indiveri}(2017)}]{Nair_Indiveri17}%
  \BibitemOpen
  \bibfield  {author} {\bibinfo {author} {\bibfnamefont {M.~V.}\ \bibnamefont
  {Nair}}\ and\ \bibinfo {author} {\bibfnamefont {G.}~\bibnamefont
  {Indiveri}},\ }\href@noop {} {\enquote {\bibinfo {title} {A differential
  memristive current-mode circuit},}\ }\bibinfo {howpublished} {European patent
  application EP 17183461.7} (\bibinfo {year} {2017}),\ \bibinfo {note} {filed
  27.07.2017}\BibitemShut {NoStop}%
\bibitem [{\citenamefont {Payvand}, \citenamefont {Muller},\ and\ \citenamefont
  {Indiveri}(2018)}]{Payvand_etal18}%
  \BibitemOpen
  \bibfield  {author} {\bibinfo {author} {\bibfnamefont {M.}~\bibnamefont
  {Payvand}}, \bibinfo {author} {\bibfnamefont {L.~K.}\ \bibnamefont {Muller}},
  \ and\ \bibinfo {author} {\bibfnamefont {G.}~\bibnamefont {Indiveri}},\
  }\bibfield  {title} {\enquote {\bibinfo {title} {Event-based circuits for
  controlling stochastic learning with memristive devices in neuromorphic
  architectures},}\ }in\ \href@noop {} {\emph {\bibinfo {booktitle} {Circuits
  and Systems (ISCAS), 2018 IEEE International Symposium on}}}\ (\bibinfo
  {organization} {IEEE},\ \bibinfo {year} {2018})\ pp.\ \bibinfo {pages}
  {1--5}\BibitemShut {NoStop}%
\bibitem [{\citenamefont {Neftci}\ \emph {et~al.}(2016)\citenamefont {Neftci},
  \citenamefont {Pedroni}, \citenamefont {Joshi}, \citenamefont {Al-Shedivat},\
  and\ \citenamefont {Cauwenberghs}}]{Neftci_etal16}%
  \BibitemOpen
  \bibfield  {author} {\bibinfo {author} {\bibfnamefont {E.~O.}\ \bibnamefont
  {Neftci}}, \bibinfo {author} {\bibfnamefont {B.~U.}\ \bibnamefont {Pedroni}},
  \bibinfo {author} {\bibfnamefont {S.}~\bibnamefont {Joshi}}, \bibinfo
  {author} {\bibfnamefont {M.}~\bibnamefont {Al-Shedivat}}, \ and\ \bibinfo
  {author} {\bibfnamefont {G.}~\bibnamefont {Cauwenberghs}},\ }\bibfield
  {title} {\enquote {\bibinfo {title} {Stochastic synapses enable efficient
  brain-inspired learning machines},}\ }\href@noop {} {\bibfield  {journal}
  {\bibinfo  {journal} {Frontiers in Neuroscience}\ }\textbf {\bibinfo {volume}
  {10}},\ \bibinfo {pages} {241} (\bibinfo {year} {2016})}\BibitemShut
  {NoStop}%
\bibitem [{\citenamefont {Brivio}\ \emph {et~al.}(2019)\citenamefont {Brivio},
  \citenamefont {Conti}, \citenamefont {Nair}, \citenamefont {Frascaroli},
  \citenamefont {Covi}, \citenamefont {Ricciardi}, \citenamefont {Indiveri},\
  and\ \citenamefont {Spiga}}]{Brivio_etal19}%
  \BibitemOpen
  \bibfield  {author} {\bibinfo {author} {\bibfnamefont {S.}~\bibnamefont
  {Brivio}}, \bibinfo {author} {\bibfnamefont {D.}~\bibnamefont {Conti}},
  \bibinfo {author} {\bibfnamefont {M.~V.}\ \bibnamefont {Nair}}, \bibinfo
  {author} {\bibfnamefont {J.}~\bibnamefont {Frascaroli}}, \bibinfo {author}
  {\bibfnamefont {E.}~\bibnamefont {Covi}}, \bibinfo {author} {\bibfnamefont
  {C.}~\bibnamefont {Ricciardi}}, \bibinfo {author} {\bibfnamefont
  {G.}~\bibnamefont {Indiveri}}, \ and\ \bibinfo {author} {\bibfnamefont
  {S.}~\bibnamefont {Spiga}},\ }\bibfield  {title} {\enquote {\bibinfo {title}
  {Extended memory lifetime in spiking neural networks employing memristive
  synapses with nonlinear conductance dynamics},}\ }\href
  {http://stacks.iop.org/0957-4484/30/i=1/a=015102} {\bibfield  {journal}
  {\bibinfo  {journal} {Nanotechnology}\ }\textbf {\bibinfo {volume} {30}},\
  \bibinfo {pages} {015102} (\bibinfo {year} {2019})}\BibitemShut {NoStop}%
\bibitem [{\citenamefont {Diederich}\ \emph {et~al.}(2018)\citenamefont
  {Diederich}, \citenamefont {Bartsch}, \citenamefont {Kohlstedt},\ and\
  \citenamefont {Ziegler}}]{Diederich_etal18}%
  \BibitemOpen
  \bibfield  {author} {\bibinfo {author} {\bibfnamefont {N.}~\bibnamefont
  {Diederich}}, \bibinfo {author} {\bibfnamefont {T.}~\bibnamefont {Bartsch}},
  \bibinfo {author} {\bibfnamefont {H.}~\bibnamefont {Kohlstedt}}, \ and\
  \bibinfo {author} {\bibfnamefont {M.}~\bibnamefont {Ziegler}},\ }\bibfield
  {title} {\enquote {\bibinfo {title} {A memristive plasticity model of
  voltage-based stdp suitable for recurrent bidirectional neural networks in
  the hippocampus},}\ }\href@noop {} {\bibfield  {journal} {\bibinfo  {journal}
  {Scientific Reports (Nature Publisher Group)}\ }\textbf {\bibinfo {volume}
  {8}},\ \bibinfo {pages} {1--12} (\bibinfo {year} {2018})}\BibitemShut
  {NoStop}%
\bibitem [{\citenamefont {Fantini}\ \emph {et~al.}(2013)\citenamefont
  {Fantini}, \citenamefont {Goux}, \citenamefont {Degraeve}, \citenamefont
  {Wouters}, \citenamefont {Raghavan}, \citenamefont {Kar}, \citenamefont
  {Belmonte}, \citenamefont {Chen}, \citenamefont {Govoreanu},\ and\
  \citenamefont {Jurczak}}]{Fantini_etal13}%
  \BibitemOpen
  \bibfield  {author} {\bibinfo {author} {\bibfnamefont {A.}~\bibnamefont
  {Fantini}}, \bibinfo {author} {\bibfnamefont {L.}~\bibnamefont {Goux}},
  \bibinfo {author} {\bibfnamefont {R.}~\bibnamefont {Degraeve}}, \bibinfo
  {author} {\bibfnamefont {D.~J.}\ \bibnamefont {Wouters}}, \bibinfo {author}
  {\bibfnamefont {N.}~\bibnamefont {Raghavan}}, \bibinfo {author}
  {\bibfnamefont {G.}~\bibnamefont {Kar}}, \bibinfo {author} {\bibfnamefont
  {A.}~\bibnamefont {Belmonte}}, \bibinfo {author} {\bibfnamefont
  {Y.}~\bibnamefont {Chen}}, \bibinfo {author} {\bibfnamefont {B.}~\bibnamefont
  {Govoreanu}}, \ and\ \bibinfo {author} {\bibfnamefont {M.}~\bibnamefont
  {Jurczak}},\ }\bibfield  {title} {\enquote {\bibinfo {title} {Intrinsic
  switching variability in hfo 2 {RRAM}},}\ }in\ \href@noop {} {\emph {\bibinfo
  {booktitle} {2013 5th {IEEE} International Memory Workshop}}}\ (\bibinfo
  {organization} {{IEEE}},\ \bibinfo {year} {2013})\ pp.\ \bibinfo {pages}
  {30--33}\BibitemShut {NoStop}%
\bibitem [{\citenamefont {Suri}\ and\ \citenamefont
  {Parmar}(2015)}]{Suri_Parmar15}%
  \BibitemOpen
  \bibfield  {author} {\bibinfo {author} {\bibfnamefont {M.}~\bibnamefont
  {Suri}}\ and\ \bibinfo {author} {\bibfnamefont {V.}~\bibnamefont {Parmar}},\
  }\bibfield  {title} {\enquote {\bibinfo {title} {Exploiting intrinsic
  variability of filamentary resistive memory for extreme learning machine
  architectures},}\ }\href@noop {} {\bibfield  {journal} {\bibinfo  {journal}
  {{IEEE} transactions on nanotechnology}\ }\textbf {\bibinfo {volume} {14}},\
  \bibinfo {pages} {963--968} (\bibinfo {year} {2015})}\BibitemShut {NoStop}%
\bibitem [{\citenamefont {Schönhals}, \citenamefont {Waser},\ and\
  \citenamefont {Wouters}(2017)}]{Schonhals_etal17}%
  \BibitemOpen
  \bibfield  {author} {\bibinfo {author} {\bibfnamefont {A.}~\bibnamefont
  {Schönhals}}, \bibinfo {author} {\bibfnamefont {R.}~\bibnamefont {Waser}}, \
  and\ \bibinfo {author} {\bibfnamefont {D.~J.}\ \bibnamefont {Wouters}},\
  }\bibfield  {title} {\enquote {\bibinfo {title} {Improvement of {SET}
  variability in {TaOxbased} resistive {RAM} devices},}\ }\href {\doibase
  10.1088/1361-6528/aa8f89} {\bibfield  {journal} {\bibinfo  {journal}
  {Nanotechnology}\ }\textbf {\bibinfo {volume} {28}},\ \bibinfo {pages}
  {465203} (\bibinfo {year} {2017})}\BibitemShut {NoStop}%
\bibitem [{\citenamefont {Prakash}\ \emph {et~al.}(2015)\citenamefont
  {Prakash}, \citenamefont {Deleruyelle}, \citenamefont {Song}, \citenamefont
  {Bocquet},\ and\ \citenamefont {Hwang}}]{Prakash_etal15}%
  \BibitemOpen
  \bibfield  {author} {\bibinfo {author} {\bibfnamefont {A.}~\bibnamefont
  {Prakash}}, \bibinfo {author} {\bibfnamefont {D.}~\bibnamefont
  {Deleruyelle}}, \bibinfo {author} {\bibfnamefont {J.}~\bibnamefont {Song}},
  \bibinfo {author} {\bibfnamefont {M.}~\bibnamefont {Bocquet}}, \ and\
  \bibinfo {author} {\bibfnamefont {H.}~\bibnamefont {Hwang}},\ }\bibfield
  {title} {\enquote {\bibinfo {title} {Resistance controllability and
  variability improvement in a taox-based resistive memory for multilevel
  storage application},}\ }\href {\doibase 10.1063/1.4922446} {\bibfield
  {journal} {\bibinfo  {journal} {Applied Physics Letters}\ }\textbf {\bibinfo
  {volume} {106}},\ \bibinfo {pages} {233104} (\bibinfo {year}
  {2015})}\BibitemShut {NoStop}%
\bibitem [{\citenamefont {Govoreanu}\ \emph {et~al.}(2015)\citenamefont
  {Govoreanu}, \citenamefont {Crotti}, \citenamefont {Subhechha}, \citenamefont
  {Zhang}, \citenamefont {Chen}, \citenamefont {Clima}, \citenamefont
  {Paraschiv}, \citenamefont {Hody}, \citenamefont {Adelmann}, \citenamefont
  {Popovici}, \citenamefont {Richard},\ and\ \citenamefont
  {Jurczak}}]{Govoreanu_etal15}%
  \BibitemOpen
  \bibfield  {author} {\bibinfo {author} {\bibfnamefont {B.}~\bibnamefont
  {Govoreanu}}, \bibinfo {author} {\bibfnamefont {D.}~\bibnamefont {Crotti}},
  \bibinfo {author} {\bibfnamefont {S.}~\bibnamefont {Subhechha}}, \bibinfo
  {author} {\bibfnamefont {L.}~\bibnamefont {Zhang}}, \bibinfo {author}
  {\bibfnamefont {Y.~Y.}\ \bibnamefont {Chen}}, \bibinfo {author}
  {\bibfnamefont {S.}~\bibnamefont {Clima}}, \bibinfo {author} {\bibfnamefont
  {V.}~\bibnamefont {Paraschiv}}, \bibinfo {author} {\bibfnamefont
  {H.}~\bibnamefont {Hody}}, \bibinfo {author} {\bibfnamefont {C.}~\bibnamefont
  {Adelmann}}, \bibinfo {author} {\bibfnamefont {M.}~\bibnamefont {Popovici}},
  \bibinfo {author} {\bibfnamefont {O.}~\bibnamefont {Richard}}, \ and\
  \bibinfo {author} {\bibfnamefont {M.}~\bibnamefont {Jurczak}},\ }\bibfield
  {title} {\enquote {\bibinfo {title} {A-{VMCO}: A novel forming-free,
  self-rectifying, analog memory cell with low-current operation,
  nonfilamentary switching and excellent variability},}\ }in\ \href@noop {}
  {\emph {\bibinfo {booktitle} {2015 Symposium on {VLSI} Technology ({VLSI}
  Technology)}}}\ (\bibinfo {organization} {{IEEE}},\ \bibinfo {year} {2015})\
  pp.\ \bibinfo {pages} {T132--T133}\BibitemShut {NoStop}%
\bibitem [{\citenamefont {Sheng}\ \emph {et~al.}(2019)\citenamefont {Sheng},
  \citenamefont {Graves}, \citenamefont {Kumar}, \citenamefont {Li},
  \citenamefont {Buchanan}, \citenamefont {Zheng}, \citenamefont {Lam},
  \citenamefont {Li},\ and\ \citenamefont {Strachan}}]{Sheng_etal19}%
  \BibitemOpen
  \bibfield  {author} {\bibinfo {author} {\bibfnamefont {X.}~\bibnamefont
  {Sheng}}, \bibinfo {author} {\bibfnamefont {C.~E.}\ \bibnamefont {Graves}},
  \bibinfo {author} {\bibfnamefont {S.}~\bibnamefont {Kumar}}, \bibinfo
  {author} {\bibfnamefont {X.}~\bibnamefont {Li}}, \bibinfo {author}
  {\bibfnamefont {B.}~\bibnamefont {Buchanan}}, \bibinfo {author}
  {\bibfnamefont {L.}~\bibnamefont {Zheng}}, \bibinfo {author} {\bibfnamefont
  {S.}~\bibnamefont {Lam}}, \bibinfo {author} {\bibfnamefont {C.}~\bibnamefont
  {Li}}, \ and\ \bibinfo {author} {\bibfnamefont {J.~P.}\ \bibnamefont
  {Strachan}},\ }\bibfield  {title} {\enquote {\bibinfo {title}
  {Low-conductance and multilevel {CMOS}-integrated nanoscale oxide
  memristors},}\ }\href@noop {} {\bibfield  {journal} {\bibinfo  {journal}
  {Advanced Electronic Materials}\ ,\ \bibinfo {pages} {1800876}} (\bibinfo
  {year} {2019})}\BibitemShut {NoStop}%
\bibitem [{\citenamefont {Freund}\ and\ \citenamefont
  {Schapire}(1997)}]{Freund_Schapire97}%
  \BibitemOpen
  \bibfield  {author} {\bibinfo {author} {\bibfnamefont {Y.}~\bibnamefont
  {Freund}}\ and\ \bibinfo {author} {\bibfnamefont {R.~E.}\ \bibnamefont
  {Schapire}},\ }\bibfield  {title} {\enquote {\bibinfo {title} {A
  decision-theoretic generalization of on-line learning and an application to
  boosting},}\ }\href@noop {} {\bibfield  {journal} {\bibinfo  {journal}
  {Journal of computer and system sciences}\ }\textbf {\bibinfo {volume}
  {55}},\ \bibinfo {pages} {119--139} (\bibinfo {year} {1997})}\BibitemShut
  {NoStop}%
\bibitem [{\citenamefont {Jaeger}\ and\ \citenamefont
  {Haas}(2004)}]{Jaeger_Haas04}%
  \BibitemOpen
  \bibfield  {author} {\bibinfo {author} {\bibfnamefont {H.}~\bibnamefont
  {Jaeger}}\ and\ \bibinfo {author} {\bibfnamefont {H.}~\bibnamefont {Haas}},\
  }\bibfield  {title} {\enquote {\bibinfo {title} {Harnessing nonlinearity:
  Predicting chaotic systems and saving energy in wireless communication},}\
  }\href@noop {} {\bibfield  {journal} {\bibinfo  {journal} {Science}\ }\textbf
  {\bibinfo {volume} {304}},\ \bibinfo {pages} {78--80} (\bibinfo {year}
  {2004})}\BibitemShut {NoStop}%
\bibitem [{\citenamefont {Maass}, \citenamefont {Natschl{\"a}ger},\ and\
  \citenamefont {Markram}(2002)}]{Maass_etal02}%
  \BibitemOpen
  \bibfield  {author} {\bibinfo {author} {\bibfnamefont {W.}~\bibnamefont
  {Maass}}, \bibinfo {author} {\bibfnamefont {T.}~\bibnamefont
  {Natschl{\"a}ger}}, \ and\ \bibinfo {author} {\bibfnamefont {H.}~\bibnamefont
  {Markram}},\ }\bibfield  {title} {\enquote {\bibinfo {title} {Real-time
  computing without stable states: A new framework for neural computation based
  on perturbations},}\ }\href@noop {} {\bibfield  {journal} {\bibinfo
  {journal} {Neural Computation}\ }\textbf {\bibinfo {volume} {14}},\ \bibinfo
  {pages} {2531--2560} (\bibinfo {year} {2002})}\BibitemShut {NoStop}%
\bibitem [{\citenamefont {Sheik}\ \emph {et~al.}(2012)\citenamefont {Sheik},
  \citenamefont {Coath}, \citenamefont {Indiveri}, \citenamefont {Denham},
  \citenamefont {Wennekers},\ and\ \citenamefont {Chicca}}]{Sheik_etal12a}%
  \BibitemOpen
  \bibfield  {author} {\bibinfo {author} {\bibfnamefont {S.}~\bibnamefont
  {Sheik}}, \bibinfo {author} {\bibfnamefont {M.}~\bibnamefont {Coath}},
  \bibinfo {author} {\bibfnamefont {G.}~\bibnamefont {Indiveri}}, \bibinfo
  {author} {\bibfnamefont {S.}~\bibnamefont {Denham}}, \bibinfo {author}
  {\bibfnamefont {T.}~\bibnamefont {Wennekers}}, \ and\ \bibinfo {author}
  {\bibfnamefont {E.}~\bibnamefont {Chicca}},\ }\bibfield  {title} {\enquote
  {\bibinfo {title} {Emergent auditory feature tuning in a real-time
  neuromorphic {VLSI} system},}\ }\href {\doibase 10.3389/fnins.2012.00017}
  {\bibfield  {journal} {\bibinfo  {journal} {Frontiers in Neuroscience}\
  }\textbf {\bibinfo {volume} {6}} (\bibinfo {year} {2012}),\
  10.3389/fnins.2012.00017}\BibitemShut {NoStop}%
\bibitem [{\citenamefont {Richter}\ \emph {et~al.}(2015)\citenamefont
  {Richter}, \citenamefont {Reinhard}, \citenamefont {Nease}, \citenamefont
  {Steil},\ and\ \citenamefont {Chicca}}]{Richter_etal15}%
  \BibitemOpen
  \bibfield  {author} {\bibinfo {author} {\bibfnamefont {O.}~\bibnamefont
  {Richter}}, \bibinfo {author} {\bibfnamefont {R.~F.}\ \bibnamefont
  {Reinhard}}, \bibinfo {author} {\bibfnamefont {S.}~\bibnamefont {Nease}},
  \bibinfo {author} {\bibfnamefont {J.}~\bibnamefont {Steil}}, \ and\ \bibinfo
  {author} {\bibfnamefont {E.}~\bibnamefont {Chicca}},\ }\bibfield  {title}
  {\enquote {\bibinfo {title} {Device mismatch in a neuromorphic system
  implements random features for regression},}\ }in\ \href {\doibase
  10.1109/BioCAS.2015.7348416} {\emph {\bibinfo {booktitle} {2015 {IEEE}
  Biomedical Circuits and Systems Conference (Bio{CAS})}}}\ (\bibinfo
  {organization} {{IEEE}},\ \bibinfo {year} {2015})\ pp.\ \bibinfo {pages}
  {1--4}\BibitemShut {NoStop}%
\bibitem [{\citenamefont {Das}\ \emph {et~al.}(2018)\citenamefont {Das},
  \citenamefont {Pradhapan}, \citenamefont {Groenendaal}, \citenamefont
  {Adiraju}, \citenamefont {Rajan}, \citenamefont {Catthoor}, \citenamefont
  {Schaafsma}, \citenamefont {Krichmar}, \citenamefont {Dutt},\ and\
  \citenamefont {Hoof}}]{Das_etal18b}%
  \BibitemOpen
  \bibfield  {author} {\bibinfo {author} {\bibfnamefont {A.}~\bibnamefont
  {Das}}, \bibinfo {author} {\bibfnamefont {P.}~\bibnamefont {Pradhapan}},
  \bibinfo {author} {\bibfnamefont {W.}~\bibnamefont {Groenendaal}}, \bibinfo
  {author} {\bibfnamefont {P.}~\bibnamefont {Adiraju}}, \bibinfo {author}
  {\bibfnamefont {R.~T.}\ \bibnamefont {Rajan}}, \bibinfo {author}
  {\bibfnamefont {F.}~\bibnamefont {Catthoor}}, \bibinfo {author}
  {\bibfnamefont {S.}~\bibnamefont {Schaafsma}}, \bibinfo {author}
  {\bibfnamefont {J.~L.}\ \bibnamefont {Krichmar}}, \bibinfo {author}
  {\bibfnamefont {N.~D.}\ \bibnamefont {Dutt}}, \ and\ \bibinfo {author}
  {\bibfnamefont {C.~V.}\ \bibnamefont {Hoof}},\ }\bibfield  {title} {\enquote
  {\bibinfo {title} {Unsupervised heart-rate estimation in wearables with
  liquid states and a probabilistic readout},}\ }\href {\doibase
  10.1016/j.neunet.2017.12.015} {\bibfield  {journal} {\bibinfo  {journal}
  {Neural networks}\ }\textbf {\bibinfo {volume} {99}},\ \bibinfo {pages}
  {134--147} (\bibinfo {year} {2018})}\BibitemShut {NoStop}%
\bibitem [{\citenamefont {Donati}\ \emph {et~al.}(2018)\citenamefont {Donati},
  \citenamefont {Payvand}, \citenamefont {Risi}, \citenamefont {Krause},
  \citenamefont {Burelo}, \citenamefont {Dalgaty}, \citenamefont {Vianello},\
  and\ \citenamefont {Indiveri}}]{Donati_etal18b}%
  \BibitemOpen
  \bibfield  {author} {\bibinfo {author} {\bibfnamefont {E.}~\bibnamefont
  {Donati}}, \bibinfo {author} {\bibfnamefont {M.}~\bibnamefont {Payvand}},
  \bibinfo {author} {\bibfnamefont {N.}~\bibnamefont {Risi}}, \bibinfo {author}
  {\bibfnamefont {R.}~\bibnamefont {Krause}}, \bibinfo {author} {\bibfnamefont
  {K.}~\bibnamefont {Burelo}}, \bibinfo {author} {\bibfnamefont
  {T.}~\bibnamefont {Dalgaty}}, \bibinfo {author} {\bibfnamefont
  {E.}~\bibnamefont {Vianello}}, \ and\ \bibinfo {author} {\bibfnamefont
  {G.}~\bibnamefont {Indiveri}},\ }\bibfield  {title} {\enquote {\bibinfo
  {title} {Processing {EMG} signals using reservoir computing on an event-based
  neuromorphic system},}\ }in\ \href {\doibase 10.1109/BIOCAS.2018.8584674}
  {\emph {\bibinfo {booktitle} {Biomedical Circuits and Systems Conference,
  ({BioCAS}), 2018}}}\ (\bibinfo {organization} {IEEE},\ \bibinfo {year}
  {2018})\ pp.\ \bibinfo {pages} {1--4}\BibitemShut {NoStop}%
\bibitem [{\citenamefont {Bauer}, \citenamefont {Muir},\ and\ \citenamefont
  {Indiveri}(2019)}]{Bauer_etal19}%
  \BibitemOpen
  \bibfield  {author} {\bibinfo {author} {\bibfnamefont {F.}~\bibnamefont
  {Bauer}}, \bibinfo {author} {\bibfnamefont {D.}~\bibnamefont {Muir}}, \ and\
  \bibinfo {author} {\bibfnamefont {G.}~\bibnamefont {Indiveri}},\ }\bibfield
  {title} {\enquote {\bibinfo {title} {Real-time ultra-low power ecg anomaly
  detection using an event-driven neuromorphic processor},}\ }\href@noop {}
  {\bibfield  {journal} {\bibinfo  {journal} {Biomedical Circuits and Systems,
  {IEEE} Transactions on}\ } (\bibinfo {year} {2019})},\ \bibinfo {note} {(in
  press)}\BibitemShut {NoStop}%
\bibitem [{\citenamefont {Gaba}\ \emph {et~al.}(2013)\citenamefont {Gaba},
  \citenamefont {Sheridan}, \citenamefont {Zhou}, \citenamefont {Choi},\ and\
  \citenamefont {Lu}}]{Gaba_etal13}%
  \BibitemOpen
  \bibfield  {author} {\bibinfo {author} {\bibfnamefont {S.}~\bibnamefont
  {Gaba}}, \bibinfo {author} {\bibfnamefont {P.}~\bibnamefont {Sheridan}},
  \bibinfo {author} {\bibfnamefont {J.}~\bibnamefont {Zhou}}, \bibinfo {author}
  {\bibfnamefont {S.}~\bibnamefont {Choi}}, \ and\ \bibinfo {author}
  {\bibfnamefont {W.}~\bibnamefont {Lu}},\ }\bibfield  {title} {\enquote
  {\bibinfo {title} {Stochastic memristive devices for computing and
  neuromorphic applications},}\ }\href@noop {} {\bibfield  {journal} {\bibinfo
  {journal} {Nanoscale}\ }\textbf {\bibinfo {volume} {5}},\ \bibinfo {pages}
  {5872--5878} (\bibinfo {year} {2013})}\BibitemShut {NoStop}%
\bibitem [{\citenamefont {Ambrogio}\ \emph {et~al.}(2016)\citenamefont
  {Ambrogio}, \citenamefont {Balatti}, \citenamefont {Milo}, \citenamefont
  {Carboni}, \citenamefont {Wang}, \citenamefont {Calderoni}, \citenamefont
  {Ramaswamy},\ and\ \citenamefont {Ielmini}}]{Ambrogio_etal16a}%
  \BibitemOpen
  \bibfield  {author} {\bibinfo {author} {\bibfnamefont {S.}~\bibnamefont
  {Ambrogio}}, \bibinfo {author} {\bibfnamefont {S.}~\bibnamefont {Balatti}},
  \bibinfo {author} {\bibfnamefont {V.}~\bibnamefont {Milo}}, \bibinfo {author}
  {\bibfnamefont {R.}~\bibnamefont {Carboni}}, \bibinfo {author} {\bibfnamefont
  {Z.-Q.}\ \bibnamefont {Wang}}, \bibinfo {author} {\bibfnamefont
  {A.}~\bibnamefont {Calderoni}}, \bibinfo {author} {\bibfnamefont
  {N.}~\bibnamefont {Ramaswamy}}, \ and\ \bibinfo {author} {\bibfnamefont
  {D.}~\bibnamefont {Ielmini}},\ }\bibfield  {title} {\enquote {\bibinfo
  {title} {Neuromorphic learning and recognition with
  one-transistor-one-resistor synapses and bistable metal oxide {RRAM}},}\
  }\href@noop {} {\bibfield  {journal} {\bibinfo  {journal} {IEEE Transactions
  on Electron Devices}\ }\textbf {\bibinfo {volume} {63}},\ \bibinfo {pages}
  {1508--1515} (\bibinfo {year} {2016})}\BibitemShut {NoStop}%
\bibitem [{\citenamefont {Yang}, \citenamefont {Strukov},\ and\ \citenamefont
  {Stewart}(2013)}]{Yang_etal13}%
  \BibitemOpen
  \bibfield  {author} {\bibinfo {author} {\bibfnamefont {J.~J.}\ \bibnamefont
  {Yang}}, \bibinfo {author} {\bibfnamefont {D.~B.}\ \bibnamefont {Strukov}}, \
  and\ \bibinfo {author} {\bibfnamefont {D.~R.}\ \bibnamefont {Stewart}},\
  }\bibfield  {title} {\enquote {\bibinfo {title} {Memristive devices for
  computing},}\ }\href@noop {} {\bibfield  {journal} {\bibinfo  {journal}
  {Nature nanotechnology}\ }\textbf {\bibinfo {volume} {8}},\ \bibinfo {pages}
  {13--24} (\bibinfo {year} {2013})}\BibitemShut {NoStop}%
\bibitem [{\citenamefont {Al-Shedivat}\ \emph {et~al.}(2015)\citenamefont
  {Al-Shedivat}, \citenamefont {Naous}, \citenamefont {Cauwenberghs},\ and\
  \citenamefont {Salama}}]{Al-Shedivat_etal15a}%
  \BibitemOpen
  \bibfield  {author} {\bibinfo {author} {\bibfnamefont {M.}~\bibnamefont
  {Al-Shedivat}}, \bibinfo {author} {\bibfnamefont {R.}~\bibnamefont {Naous}},
  \bibinfo {author} {\bibfnamefont {G.}~\bibnamefont {Cauwenberghs}}, \ and\
  \bibinfo {author} {\bibfnamefont {K.~N.}\ \bibnamefont {Salama}},\ }\bibfield
   {title} {\enquote {\bibinfo {title} {Memristors empower spiking neurons with
  stochasticity},}\ }\href@noop {} {\bibfield  {journal} {\bibinfo  {journal}
  {IEEE Journal on Emerging and Selected Topics in Circuits and Systems}\
  }\textbf {\bibinfo {volume} {5}},\ \bibinfo {pages} {242--253} (\bibinfo
  {year} {2015})}\BibitemShut {NoStop}%
\bibitem [{\citenamefont {Suri}\ \emph {et~al.}(2013)\citenamefont {Suri},
  \citenamefont {Querlioz}, \citenamefont {Bichler}, \citenamefont {Palma},
  \citenamefont {Vianello}, \citenamefont {Vuillaume}, \citenamefont {Gamrat},\
  and\ \citenamefont {DeSalvo}}]{Suri_etal13}%
  \BibitemOpen
  \bibfield  {author} {\bibinfo {author} {\bibfnamefont {M.}~\bibnamefont
  {Suri}}, \bibinfo {author} {\bibfnamefont {D.}~\bibnamefont {Querlioz}},
  \bibinfo {author} {\bibfnamefont {O.}~\bibnamefont {Bichler}}, \bibinfo
  {author} {\bibfnamefont {G.}~\bibnamefont {Palma}}, \bibinfo {author}
  {\bibfnamefont {E.}~\bibnamefont {Vianello}}, \bibinfo {author}
  {\bibfnamefont {D.}~\bibnamefont {Vuillaume}}, \bibinfo {author}
  {\bibfnamefont {C.}~\bibnamefont {Gamrat}}, \ and\ \bibinfo {author}
  {\bibfnamefont {B.}~\bibnamefont {DeSalvo}},\ }\bibfield  {title} {\enquote
  {\bibinfo {title} {Bio-inspired stochastic computing using binary {CBRAM}
  synapses},}\ }\href@noop {} {\bibfield  {journal} {\bibinfo  {journal}
  {Electron Devices, {IEEE} Transactions on}\ }\textbf {\bibinfo {volume}
  {60}},\ \bibinfo {pages} {2402--2409} (\bibinfo {year} {2013})}\BibitemShut
  {NoStop}%
\bibitem [{\citenamefont {Balatti}\ \emph {et~al.}(2016)\citenamefont
  {Balatti}, \citenamefont {Ambrogio}, \citenamefont {Carboni}, \citenamefont
  {Milo}, \citenamefont {Wang}, \citenamefont {Calderoni}, \citenamefont
  {Ramaswamy},\ and\ \citenamefont {Ielmini}}]{Balatti_etal16}%
  \BibitemOpen
  \bibfield  {author} {\bibinfo {author} {\bibfnamefont {S.}~\bibnamefont
  {Balatti}}, \bibinfo {author} {\bibfnamefont {S.}~\bibnamefont {Ambrogio}},
  \bibinfo {author} {\bibfnamefont {R.}~\bibnamefont {Carboni}}, \bibinfo
  {author} {\bibfnamefont {V.}~\bibnamefont {Milo}}, \bibinfo {author}
  {\bibfnamefont {Z.}~\bibnamefont {Wang}}, \bibinfo {author} {\bibfnamefont
  {A.}~\bibnamefont {Calderoni}}, \bibinfo {author} {\bibfnamefont
  {N.}~\bibnamefont {Ramaswamy}}, \ and\ \bibinfo {author} {\bibfnamefont
  {D.}~\bibnamefont {Ielmini}},\ }\bibfield  {title} {\enquote {\bibinfo
  {title} {Physical unbiased generation of random numbers with coupled
  resistive switching devices},}\ }\href@noop {} {\bibfield  {journal}
  {\bibinfo  {journal} {{IEEE} Transactions on Electron Devices}\ }\textbf
  {\bibinfo {volume} {63}},\ \bibinfo {pages} {2029--2035} (\bibinfo {year}
  {2016})}\BibitemShut {NoStop}%
\bibitem [{\citenamefont {Habenschuss}, \citenamefont {Jonke},\ and\
  \citenamefont {Maass}(2013)}]{Habenschuss_etal13}%
  \BibitemOpen
  \bibfield  {author} {\bibinfo {author} {\bibfnamefont {S.}~\bibnamefont
  {Habenschuss}}, \bibinfo {author} {\bibfnamefont {Z.}~\bibnamefont {Jonke}},
  \ and\ \bibinfo {author} {\bibfnamefont {W.}~\bibnamefont {Maass}},\
  }\bibfield  {title} {\enquote {\bibinfo {title} {Stochastic computations in
  cortical microcircuit models},}\ }\href@noop {} {\bibfield  {journal}
  {\bibinfo  {journal} {PLoS computational biology}\ }\textbf {\bibinfo
  {volume} {9}},\ \bibinfo {pages} {e1003311} (\bibinfo {year}
  {2013})}\BibitemShut {NoStop}%
\bibitem [{\citenamefont {Destexhe}\ and\ \citenamefont
  {Contreras}(2006)}]{Destexhe_Contreras06}%
  \BibitemOpen
  \bibfield  {author} {\bibinfo {author} {\bibfnamefont {A.}~\bibnamefont
  {Destexhe}}\ and\ \bibinfo {author} {\bibfnamefont {D.}~\bibnamefont
  {Contreras}},\ }\bibfield  {title} {\enquote {\bibinfo {title} {Neuronal
  computations with stochastic network states},}\ }\href@noop {} {\bibfield
  {journal} {\bibinfo  {journal} {Science}\ }\textbf {\bibinfo {volume}
  {314}},\ \bibinfo {pages} {85--90} (\bibinfo {year} {2006})}\BibitemShut
  {NoStop}%
\bibitem [{\citenamefont {Fusi}\ and\ \citenamefont
  {Senn}(2006)}]{Fusi_Senn06}%
  \BibitemOpen
  \bibfield  {author} {\bibinfo {author} {\bibfnamefont {S.}~\bibnamefont
  {Fusi}}\ and\ \bibinfo {author} {\bibfnamefont {W.}~\bibnamefont {Senn}},\
  }\bibfield  {title} {\enquote {\bibinfo {title} {Eluding oblivion with smart
  stochastic selection of synaptic updates},}\ }\href@noop {} {\bibfield
  {journal} {\bibinfo  {journal} {Chaos, An Interdisciplinary Journal of
  Nonlinear Science}\ }\textbf {\bibinfo {volume} {16}},\ \bibinfo {pages}
  {1--11} (\bibinfo {year} {2006})}\BibitemShut {NoStop}%
\bibitem [{\citenamefont {Ginzburg}\ and\ \citenamefont
  {Sompolinsky}(1994)}]{Ginzburg_Sompolinsky94}%
  \BibitemOpen
  \bibfield  {author} {\bibinfo {author} {\bibfnamefont {I.}~\bibnamefont
  {Ginzburg}}\ and\ \bibinfo {author} {\bibfnamefont {H.}~\bibnamefont
  {Sompolinsky}},\ }\bibfield  {title} {\enquote {\bibinfo {title} {Theory of
  correlations in stochastic neural networks},}\ }\href@noop {} {\bibfield
  {journal} {\bibinfo  {journal} {Physical Review E}\ }\textbf {\bibinfo
  {volume} {50}},\ \bibinfo {pages} {3171--91} (\bibinfo {year}
  {1994})}\BibitemShut {NoStop}%
\bibitem [{\citenamefont {Dalgaty}\ \emph {et~al.}(2019)\citenamefont
  {Dalgaty}, \citenamefont {Payvand}, \citenamefont {De~Salvo}, \citenamefont
  {Casaz}, \citenamefont {Lama}, \citenamefont {Nowak}, \citenamefont
  {Indiveri},\ and\ \citenamefont {Vianello}}]{Dalgaty_etal19}%
  \BibitemOpen
  \bibfield  {author} {\bibinfo {author} {\bibfnamefont {T.}~\bibnamefont
  {Dalgaty}}, \bibinfo {author} {\bibfnamefont {M.}~\bibnamefont {Payvand}},
  \bibinfo {author} {\bibfnamefont {B.}~\bibnamefont {De~Salvo}}, \bibinfo
  {author} {\bibfnamefont {J.}~\bibnamefont {Casaz}}, \bibinfo {author}
  {\bibfnamefont {G.}~\bibnamefont {Lama}}, \bibinfo {author} {\bibfnamefont
  {E.}~\bibnamefont {Nowak}}, \bibinfo {author} {\bibfnamefont
  {G.}~\bibnamefont {Indiveri}}, \ and\ \bibinfo {author} {\bibfnamefont
  {E.}~\bibnamefont {Vianello}},\ }\bibfield  {title} {\enquote {\bibinfo
  {title} {Hybrid cmos-rram neurons with intrinsic plasticit},}\ }in\
  \href@noop {} {\emph {\bibinfo {booktitle} {International Symposium on
  Circuits and Systems ({ISCAS}), 2019}}}\ (\bibinfo {organization} {IEEE},\
  \bibinfo {year} {2019})\BibitemShut {NoStop}%
\bibitem [{\citenamefont {Yousefzadeh}\ \emph {et~al.}(2018)\citenamefont
  {Yousefzadeh}, \citenamefont {Stromatias}, \citenamefont {Soto},
  \citenamefont {Serrano-Gotarredona},\ and\ \citenamefont
  {Linares-Barranco}}]{Yousefzadeh_etal18}%
  \BibitemOpen
  \bibfield  {author} {\bibinfo {author} {\bibfnamefont {A.}~\bibnamefont
  {Yousefzadeh}}, \bibinfo {author} {\bibfnamefont {E.}~\bibnamefont
  {Stromatias}}, \bibinfo {author} {\bibfnamefont {M.}~\bibnamefont {Soto}},
  \bibinfo {author} {\bibfnamefont {T.}~\bibnamefont {Serrano-Gotarredona}}, \
  and\ \bibinfo {author} {\bibfnamefont {B.}~\bibnamefont {Linares-Barranco}},\
  }\bibfield  {title} {\enquote {\bibinfo {title} {On practical issues for
  stochastic stdp hardware with 1-bit synaptic weights},}\ }\href@noop {}
  {\bibfield  {journal} {\bibinfo  {journal} {Frontiers in neuroscience}\
  }\textbf {\bibinfo {volume} {12}} (\bibinfo {year} {2018})}\BibitemShut
  {NoStop}%
\bibitem [{\citenamefont {Leugering}\ and\ \citenamefont
  {Pipa}(2018)}]{Leugering_Pipa18}%
  \BibitemOpen
  \bibfield  {author} {\bibinfo {author} {\bibfnamefont {J.}~\bibnamefont
  {Leugering}}\ and\ \bibinfo {author} {\bibfnamefont {G.}~\bibnamefont
  {Pipa}},\ }\bibfield  {title} {\enquote {\bibinfo {title} {A unifying
  framework of synaptic and intrinsic plasticity in neural populations},}\
  }\href@noop {} {\bibfield  {journal} {\bibinfo  {journal} {Neural
  computation}\ }\textbf {\bibinfo {volume} {30}},\ \bibinfo {pages} {945--986}
  (\bibinfo {year} {2018})}\BibitemShut {NoStop}%
\bibitem [{\citenamefont {Fusi}\ and\ \citenamefont
  {Abbott}(2007)}]{Fusi_Abbott07}%
  \BibitemOpen
  \bibfield  {author} {\bibinfo {author} {\bibfnamefont {S.}~\bibnamefont
  {Fusi}}\ and\ \bibinfo {author} {\bibfnamefont {L.}~\bibnamefont {Abbott}},\
  }\bibfield  {title} {\enquote {\bibinfo {title} {Limits on the memory storage
  capacity of bounded synapses},}\ }\href@noop {} {\bibfield  {journal}
  {\bibinfo  {journal} {Nature Neuroscience}\ }\textbf {\bibinfo {volume}
  {10}},\ \bibinfo {pages} {485--493} (\bibinfo {year} {2007})}\BibitemShut
  {NoStop}%
\bibitem [{\citenamefont {Ganguli}, \citenamefont {Huh},\ and\ \citenamefont
  {Sompolinsky}(2008)}]{Ganguli_etal08}%
  \BibitemOpen
  \bibfield  {author} {\bibinfo {author} {\bibfnamefont {S.}~\bibnamefont
  {Ganguli}}, \bibinfo {author} {\bibfnamefont {D.}~\bibnamefont {Huh}}, \ and\
  \bibinfo {author} {\bibfnamefont {H.}~\bibnamefont {Sompolinsky}},\
  }\bibfield  {title} {\enquote {\bibinfo {title} {Memory traces in dynamical
  systems},}\ }\href@noop {} {\bibfield  {journal} {\bibinfo  {journal}
  {Proceedings of the National Academy of Sciences}\ }\textbf {\bibinfo
  {volume} {105}},\ \bibinfo {pages} {18970--18975} (\bibinfo {year}
  {2008})}\BibitemShut {NoStop}%
\bibitem [{\citenamefont {Frascaroli}\ \emph {et~al.}(2018)\citenamefont
  {Frascaroli}, \citenamefont {Brivio}, \citenamefont {Covi},\ and\
  \citenamefont {Spiga}}]{Frascaroli_etal18}%
  \BibitemOpen
  \bibfield  {author} {\bibinfo {author} {\bibfnamefont {J.}~\bibnamefont
  {Frascaroli}}, \bibinfo {author} {\bibfnamefont {S.}~\bibnamefont {Brivio}},
  \bibinfo {author} {\bibfnamefont {E.}~\bibnamefont {Covi}}, \ and\ \bibinfo
  {author} {\bibfnamefont {S.}~\bibnamefont {Spiga}},\ }\bibfield  {title}
  {\enquote {\bibinfo {title} {Evidence of soft bound behaviour in analogue
  memristive devices for neuromorphic computing},}\ }\href@noop {} {\bibfield
  {journal} {\bibinfo  {journal} {Scientific reports}\ }\textbf {\bibinfo
  {volume} {8}},\ \bibinfo {pages} {71--78} (\bibinfo {year}
  {2018})}\BibitemShut {NoStop}%
\bibitem [{\citenamefont {{Del Giudice}}, \citenamefont {Fusi},\ and\
  \citenamefont {Mattia}(2003)}]{Del-Giudice_etal03}%
  \BibitemOpen
  \bibfield  {author} {\bibinfo {author} {\bibfnamefont {P.}~\bibnamefont {{Del
  Giudice}}}, \bibinfo {author} {\bibfnamefont {S.}~\bibnamefont {Fusi}}, \
  and\ \bibinfo {author} {\bibfnamefont {M.}~\bibnamefont {Mattia}},\
  }\bibfield  {title} {\enquote {\bibinfo {title} {Modeling the formation of
  working memory with networks of integrate-and-fire neurons connected by
  plastic synapses},}\ }\href@noop {} {\bibfield  {journal} {\bibinfo
  {journal} {Journal of Physiology Paris 97}\ ,\ \bibinfo {pages} {659--681}}
  (\bibinfo {year} {2003})}\BibitemShut {NoStop}%
\bibitem [{\citenamefont {Zenke}, \citenamefont {Agnes},\ and\ \citenamefont
  {Gerstner}(2015)}]{Zenke_etal15}%
  \BibitemOpen
  \bibfield  {author} {\bibinfo {author} {\bibfnamefont {F.}~\bibnamefont
  {Zenke}}, \bibinfo {author} {\bibfnamefont {E.~J.}\ \bibnamefont {Agnes}}, \
  and\ \bibinfo {author} {\bibfnamefont {W.}~\bibnamefont {Gerstner}},\
  }\bibfield  {title} {\enquote {\bibinfo {title} {Diverse synaptic plasticity
  mechanisms orchestrated to form and retrieve memories in spiking neural
  networks},}\ }\href {\doibase 10.1038/ncomms7922} {\bibfield  {journal}
  {\bibinfo  {journal} {Nature Communications}\ }\textbf {\bibinfo {volume}
  {6}},\ \bibinfo {pages} {1--13} (\bibinfo {year} {2015})}\BibitemShut
  {NoStop}%
\bibitem [{\citenamefont {Bill}\ and\ \citenamefont
  {Legenstein}(2014)}]{Bill_Legenstein14}%
  \BibitemOpen
  \bibfield  {author} {\bibinfo {author} {\bibfnamefont {J.}~\bibnamefont
  {Bill}}\ and\ \bibinfo {author} {\bibfnamefont {R.}~\bibnamefont
  {Legenstein}},\ }\bibfield  {title} {\enquote {\bibinfo {title} {A compound
  memristive synapse model for statistical learning through stdp in spiking
  neural networks},}\ }\href@noop {} {\bibfield  {journal} {\bibinfo  {journal}
  {Frontiers in neuroscience}\ }\textbf {\bibinfo {volume} {8}},\ \bibinfo
  {pages} {1--18} (\bibinfo {year} {2014})}\BibitemShut {NoStop}%
\bibitem [{\citenamefont {Plastiras}\ \emph {et~al.}(2018)\citenamefont
  {Plastiras}, \citenamefont {Terzi}, \citenamefont {Kyrkou},\ and\
  \citenamefont {Theocharidcs}}]{Plastiras_etal18}%
  \BibitemOpen
  \bibfield  {author} {\bibinfo {author} {\bibfnamefont {G.}~\bibnamefont
  {Plastiras}}, \bibinfo {author} {\bibfnamefont {M.}~\bibnamefont {Terzi}},
  \bibinfo {author} {\bibfnamefont {C.}~\bibnamefont {Kyrkou}}, \ and\ \bibinfo
  {author} {\bibfnamefont {T.}~\bibnamefont {Theocharidcs}},\ }\bibfield
  {title} {\enquote {\bibinfo {title} {Edge intelligence: Challenges and
  opportunities of near-sensor machine learning applications},}\ }in\
  \href@noop {} {\emph {\bibinfo {booktitle} {2018 IEEE 29th International
  Conference on Application-specific Systems, Architectures and Processors
  (ASAP)}}}\ (\bibinfo {organization} {IEEE},\ \bibinfo {year} {2018})\ pp.\
  \bibinfo {pages} {1--7}\BibitemShut {NoStop}%
\end{thebibliography}%

\end{document}